\documentclass[a4paper,11pt]{article}

\usepackage[english]{babel}

\usepackage[latin1]{inputenc}
\usepackage[toc,page]{appendix}
\usepackage{amsfonts,amsbsy,bm,euscript,mathrsfs}
\usepackage{amssymb,stmaryrd,faktor,slashed}
\usepackage{color}

\usepackage[nosort]{cite}

\usepackage{authblk, physics, multirow, comment}

\usepackage{graphicx}
\usepackage[verbose]{wrapfig}
\usepackage{caption}
\usepackage{float}

\numberwithin{equation}{section}

\usepackage{enumitem}

\newcommand{\ie}{{\it i.e.\ }}
\newcommand{\cf}{{\it c.f.\ }}

\usepackage{jheppub}

\arxivnumber{2508.00987}

\title{\boldmath \Large\bf Interplay of
Intersecting Branes and Consistent Embeddings}

\author[a]{Jieming Lin\,,}
\author[b]{Torben Skrzypek}
\author[a]{and K. S. Stelle}
\affiliation[a]{Abdus Salam Centre for Theoretical Physics, Imperial College London,\\ Prince Consort Road, London, SW7 2AZ, UK}
\affiliation[b]{Deutsches Elektronen-Synchrotron DESY, \\ Notkestra{\ss}e 85, 22607 Hamburg, Germany}

\emailAdd{jieming.lin22@imperial.ac.uk, torben.skrzypek@desy.de}

\abstract{We explore consistent truncations from 10d/11d supergravity to supergravity theories on brane worldvolumes. Supersymmetric black-brane solutions to these lower-dimensional theories can be uplifted to 10d/11d supergravity and interpreted as intersecting-brane solutions with a particular smearing pattern. We survey this novel family of intersecting-brane solutions which also includes cases that are not related to previously known consistent truncations. Turning the argument on its head, we argue that the knowledge of such solutions allows us to generate appropriate embedding ans\"atze for consistent truncations. We demonstrate this by deriving new consistent embeddings of pure 5d $\mathcal{N}=4$ supergravity on D4-branes in type IIA theory and of pure 6d $\mathcal{N}=(1,1)$ supergravity on NS5-branes in type IIB theory. }

\begin{document}

\begin{flushright}\small{Imperial/TP/2025/KS/{02}}\\ \small{DESY-25-106}\end{flushright}
\vspace{0.1cm}

\maketitle
\flushbottom

\section{Introduction}\label{intro}
A characteristic feature of superstring and supergravity theories is a higher-dimensional spacetime as the original home of the panoply of solutions: $d=11$ for M-Theory or $d=10$ for maximal string theories or supergravities. In proposals for how to relate such solutions to more practical physics in $d=4$, the most common approach is to consider solutions in which the excess dimensions are compact. A different interesting family of solutions is that of branes: spacetimes with flat or asymptotically flat sub-spacetimes which can be considered the arena for lower-dimensional physics, but with non-compact transverse spaces filling the remaining dimensions back up to 10d/11d. Further identifying branes as boundaries of open strings, one can naturally embed gauge theories on the brane worldvolume, paving a way towards an embedding of the standard model. Moreover, the celebrated AdS/CFT duality was discovered in this context \cite{Maldacena:1997re,Witten:1998qj}.

A related family of supergravity solutions which involves multiple brane structures of various dimensions is the general family of ``intersecting branes''.\footnote{Considering open strings on such brane intersections allows for the introduction of charged matter content in the worldvolume gauge theory. Here, however, we focus on supergravity aspects. } A by no means complete set of references for the study of such solutions is
\cite{Tseytlin:1996bh,Gauntlett:1996pb,Behrndt:1996pm,Tseytlin:1996zb,Bergshoeff:1996rn,Tseytlin:1997cs,Gauntlett:1997cv,Youm:1999ti,Smith:2002wn}. An important feature of these solutions is summarised in a ``harmonic function rule'' \cite{Tseytlin:1996bh,Gauntlett:1996pb}, where the contribution of each brane is encoded in a harmonic function $H_i$, depending on some or all of the coordinates transverse to its worldvolume. For example, the metric of two intersecting branes, extended in dimensions $(x,y_1)$ and $(x,y_2)$, respectively, takes the schematic form
\begin{equation}\label{eq:schematic}
  \dd s^2= H_1^{\alpha_1} H_2^{\alpha_2}\dd x^2 + H_1^{\alpha_1} H_2^{\beta_2}\dd y_1^2 + H_1^{\beta_1} H_2^{\alpha2}\dd y_2^2 + H_1^{\beta_1} H_2^{\beta_2}\dd z^2 \,,
\end{equation}
where the coefficients $\alpha_i, \beta_i$ and the dimensionalities of the various subspaces depend on the specific setup. An additional aspect of these solutions concerns the domains of dependence of the various harmonic functions. For the original type of intersecting-brane solutions, the harmonic functions $H_i(z)$ did not depend on ``relative transverse dimensions", \ie spacetime dimensions shared partially as a worldvolume dimension of one component while also appearing as a transverse dimension of another ($y_1$ and $y_2$ in \eqref{eq:schematic}). This ``smearing'' of functional dependence on relative transverse coordinates is analogous to the smearing imposed on brane transverse dimensions made prior to ``vertical dimensional reduction'' \cite{Lu:1996mg}, although here the intention is not to actually carry out a dimensional reduction but is rather to build up a complex higher-dimensional spacetime from a variety of component branes of diverse worldvolume structures. 

In this paper we consider a different family of intersecting branes and its relation to the framework of ``consistent truncations''.\footnote{ Since in this paper, we will be concerned with the relation of lower-dimensional supergravity solutions to a higher-dimensional family of intersecting branes, we will mostly prefer the term ``embedding'' to ``truncation''.} These are ans\"atze for specific families of higher-dimensional solutions based on an underlying supersymmetric ``skeleton'' brane on whose worldvolume one may consistently embed a lower-dimensional supergravity. Consistency in such a context means that an {\em arbitrary} solution of the skeleton worldvolume sub-theory may be reinterpreted, or ``lifted'' to a fully valid solution of the host higher-dimensional theory. Such consistent truncations to pure supergravity theories on brane worldvolumes have been discussed in Refs \cite{Lu:2000xc,Leung:2022nhy} and we will further expand the catalogue of available truncation scenarios in this paper. Let us emphasise that the benefit of such consistent truncations is that they capture realistic lower-dimensional phenomena, even highly non-perturbative objects such as black holes,  but are still supported by a stringy UV completion in higher dimensions.

Inspired by such black holes solutions, we investigate the uplift of some {\em particular} supersymmetric solutions in the lower-dimensional supergravities, which may be interpreted as a novel type of $\tfrac{1}{4}$-BPS intersecting-brane solutions in the higher-dimensional host theory. This type of intersecting-brane solutions can also be considered ab initio in the higher dimension. They still obey a harmonic function rule analogous to \eqref{eq:schematic} but involve three branes and feature a particular coordinate dependence for the various harmonic functions. The skeleton brane plays a special r\^ole, being ``fully localised''\footnote{In \eqref{eq:schematic}, ``full localisation" would correspond to a harmonic function $H_1(y_2,z)$ depending on all its transverse coordinates.} while the non-skeleton components contribute harmonic functions dependent only on certain skeleton-brane worldvolume coordinates while they are smeared in the transverse directions of the skeleton brane. This gives an overall set of harmonic functional dependences that is distinct from the relative-transverse-smeared solutions of Refs\,\cite{Tseytlin:1996bh,Gauntlett:1996pb,Behrndt:1996pm,Tseytlin:1996zb,Bergshoeff:1996rn,Tseytlin:1997cs,Gauntlett:1997cv,Youm:1999ti,Smith:2002wn}. From the point of view of the consistent embedding on the skeleton brane, we may identify these intersections with supersymmetric black-brane or supersymmetric black-hole solutions. This identification of extremal black holes as intersecting-brane solutions is reminiscent of the famous black-hole microstate analysis in \cite{Strominger:1996sh}, but has the potential to extend to less supersymmetric scenarios, which would allow for a string-theory embedding of more realistic black-hole solutions.
 
In Section \ref{Intersecting brane}, after a quick review of the traditional harmonic function rule \cite{Tseytlin:1996bh,Gauntlett:1996pb}, we present the new family of intersecting-brane solutions with a fully-localised component. These have a metric structure consistent with the usual harmonic function rule \cite{Tseytlin:1996bh,Gauntlett:1996pb}, but the functional dependence of the component harmonic functions is different from that of the previous intersecting-brane solutions, and an additional harmonic function makes its appearance in the form-fields. One of the component branes will be fully localised, with unsmeared coordinate dependence in its full transverse space, while the other components are mutually smeared. We discuss which combinations of three branes preserve one quarter of the supersymmetry and note that for each such combination, any one of the component branes may be chosen to be the fully localised one. We thus generate a large catalogue of new intersecting-brane solutions. 

In Section \ref{Skeleton}, we reinterpret these intersecting-brane solutions from the point of view of the worldvolume theory of the fully-localised component brane. This interpretation relies on the existence of a consistent embedding of a lower-dimensional supergravity on the worldvolume of the fully-localised brane, now viewed as a ``skeleton'' brane in the language of Ref.\,\cite{Leung:2022nhy}. Some examples are provided in which the worldvolume embedded solution is a black hole or a black string. This duality of viewpoints -- either as members of the new class of intersecting branes with a fully-localised component, or as consistently embedded worldvolume supergravity solutions -- also motivates a method for inferring the detailed structure of a consistent embedding ansatz on the skeleton worldvolume, starting from the host higher-dimensional theory.

In Section \ref{Embedding} we make use of the previously uncovered interplay of intersecting branes and consistent embeddings to construct new consistent embeddings on brane worldvolumes. This is followed by a conclusion and an outlook in Section \ref{outlook}. In Appendix \ref{app:conventions}, we present our conventions, details of 10d type IIA/B, 6d $\mathcal{N}=(1,1)$ and 5d $N=4$ supergravities, followed in Appendix \ref{app:susy} by the supersymmetry projections of the considered BPS branes. In Appendix \ref{app:fermion-consistency}, we extend the consistent-embedding analysis to include details of the fermionic sector, giving details of the fermionic ans\"atze and proofs of consistency.

\

\subsubsection*{Note added} 
Kellogg S. Stelle passed away in October 2025, after this paper had been submitted. The final version has been prepared for publication by J. Lin and T. Skrzypek.\\
Kelly was a valued colleague, friend and supervisor. He made important and lasting contributions to physics and the community, and will be deeply missed. 
\begin{flushright}
---JL and TS
\end{flushright}

\section{Intersecting-brane solutions with a fully localised component}\label{Intersecting brane}
In 10d type II supergravities, there are known flat brane solutions preserving half the supersymmetry. The metric and the dilaton of a D$p$-brane in the Einstein frame\footnote{The string frame is related via $\dd \hat{s}_\text{str}^2=e^{\hat{\Phi}/2}\dd \hat{s}_\text{E}^2$.} are given by
\begin{equation}
  \dd \hat{s}^2_{10} = H(y)^{\frac{p-7}{8}}\eta_{\mu\nu}\,\dd x^\mu \dd x^\nu + H(y)^{\frac{p+7}{8}}\delta_{IJ}\,\dd y^I\dd y^J\,,\qquad e^{\hat{\Phi}}=H(y)^{\frac{3-p}{4}}\,,
  \label{eq:p-brane solution}
\end{equation}
while the electric or magnetic flux sourced by the brane is 
\begin{equation}
  \hat{F}^\text{ele}_{(p+2)} = \partial_I H^{-1}\, \dd x^0\wedge \cdots \wedge \dd x^p \wedge \dd y^I\,,\quad \hat{F}^{\text{mag}}_{(8-p)} = \frac{1}{(8-p)!} \varepsilon_{I_1\cdots I_{9-p}}\partial_{I_1}H \,\dd y^{I_2}\wedge\dots\wedge\dd y^{I_{9-p}}\,,
\end{equation}
respectively. The function $H(y)$ is a harmonic function in the flat underlying transverse space, \ie 
\begin{equation}
  \partial_I^2H(y)=0\,.
  \label{eq:harmonic function}
\end{equation}
A classical solution describing a stack of $N$ D$p$-branes sitting at transverse radius $r=0$ is given by 
\begin{equation}
  H(y) = 1 + \frac{\alpha N}{r^{7-p}}\,,
\end{equation} 
where $\alpha$ is an appropriate constant with dimension $[\text{Length}]^{7-p}$. 
Similar backgrounds are also available for the F1-string, the NS5-brane \cite{Blumenhagen:2013fgp} as well as for the M-branes in 11d supergravity \cite{Duff:1990xz, Gueven:1992hh}.

Intersecting-brane solutions can be constructed using the established harmonic function rule \cite{Tseytlin:1996bh, Gauntlett:1996pb}. The spacetime metric and the exponential of the dilaton are obtained by multiplying the corresponding expressions for each constituent brane given in \eqref{eq:p-brane solution}, while the fluxes are determined by summing the individual contributions from each brane. In this class of solutions, the harmonic functions depend only on the overall transverse space. Physically, these branes are considered as smeared across the relative transverse directions analogously to the smeared transverse-space dependence of brane solutions prepared for vertical dimensional reduction \cite{Lu:1996mg}. For example, the F1-D5 brane solution given by the harmonic function rule reads
\begin{equation}\label{eq:harmonic}
  \begin{aligned}
    &\dd \hat{s}^2_{10} = -H_1^{-\frac34}H_2^{-\frac14}\dd t^2 + H_1^{-\frac34}H_2^{\frac34}\dd x_1^2 + H_1^{\frac14}H_2^{-\frac14}(\dd x_2^2 + \cdots + \dd x_6^2)+ H_1^{\frac14}H_2^{\frac34}(\dd y_1^2 + \dd y_2^2 + \dd y_3^2)\,,\\
    &\hat{H}_{(3)} = H_1^{-2}\partial_I H_1 \dd t\wedge \dd x_1\wedge \dd y^I\,,\quad \tilde{F}_{(3)}= \frac12 \varepsilon_{IJK}\partial_I H_2 \, \dd x_1\wedge \dd y^J\wedge \dd y^K\,,\quad e^{\hat{\Phi}} = H_1^{-\frac12}H_2^{-\frac12}\,.\\
  \end{aligned}
\end{equation}
Here, $H_1$ and $H_2$ are two independent harmonic functions of the overall transverse coordinates $y^I$ only. They are associated to the F1-string and the D5-brane, respectively. Explicitly, we have the conditions
\begin{equation}
  \partial_I^2H_{1}(y)=\partial_I^2H_{2}(y)=0\,.
\end{equation}

In principle, fully localised intersecting-brane solutions may also exist. In that case, each harmonic function would then depend on all transverse coordinates of the associated brane and should solve a corresponding curved-space Laplace equation, the direct curved space generalisation of \eqref{eq:harmonic function}. However, solving these Laplace equations in a coupled system of multiple branes is analytically challenging. To our knowledge, only partially localised solutions have been constructed in a near-horizon region \cite{Youm:1999ti}. In that class of solutions, one brane is fully localised, while the other is smeared over the relative transverse space of the localised brane. The curved-space Laplace equation is then solved in the near-horizon limit of the localised brane.

The mutually smeared brane solutions \eqref{eq:harmonic} can be directly generalised to higher numbers of intersecting branes. For example, a case of three intersecting M2-branes was discussed in Refs \cite{Tseytlin:1996bh,Gauntlett:1996pb}. In general, such $n$-brane solutions will preserve $1/2^n$ supersymmetry. However, supersymmetry can be enhanced in cases where the Killing-spinor projections of the respective branes are not independent. A well-known example \cite{Gauntlett:1997cv} is a configuration of two M5-branes and one M2-brane sharing one spatial direction. The Killing-spinor projections of the two M5-branes are 
\begin{equation}
  \Gamma_{\underline{012345}}\varepsilon = c_1 \varepsilon\,,\qquad \Gamma_{\underline{016789}}\varepsilon = c_2 \varepsilon\,,
  \label{eq:M5-brane-Projection}
\end{equation}
where $c_1,\,c_2=\pm1$ are two independent constants labelling the orientations of the M5-branes. Multiplying these two conditions, one sees that they immediately imply the following projection condition
\begin{equation}
  \Gamma_{\underline{01,10}}\varepsilon = c_1 c_2\varepsilon\,,
  \label{eq:M2-brane-Projection}
\end{equation}
reminiscent of an M2-brane Killing-spinor condition.
Hence, if one adds an M2-brane with the appropriate orientation, the solution still preserves $\tfrac14$ of the supersymmetry instead of the naively expected $\tfrac18$. One may thus add this brane ``for free" at no cost to the supersymmetry of the system. However, if one instead adds an M2-brane with the opposite orientation, all supersymmetry is broken. Further discussions of such intersecting-brane solutions can be found in the reviews \cite{Gauntlett:1997cv, Smith:2002wn}.

\

In this paper, we combine the aspects of partial localisation and ``free" additional branes. We accordingly find a new class of partially localised brane intersections involving three branes that satisfy a generalised harmonic function rule and that preserve eight supercharges, \ie they are $\tfrac{1}{4}$-BPS solutions. Starting from a particular smeared brane intersection of two branes, we can introduce a third brane, fully localised, whose worldvolume spans the intersection space and the overall transverse space of the smeared-brane pair. The smeared brane pair may then be reinterpreted as the 10d/11d uplift of black-brane solutions in the worldvolume supergravity of the fully localised brane \cite{Leung:2022nhy}. This perspective will be discussed in Section \ref{Skeleton}. 

\begin{table}[ht]
\centering
\begin{tabular}{|ccc|ccc|ccc|}
\hline
\multicolumn{3}{|c|}{\textbf{IIA}}                    & \multicolumn{3}{c|}{\textbf{IIB}}                   & \multicolumn{1}{c|}{\textbf{11d}}\\ \hline
\multicolumn{1}{|c|}{F1-D0-D8}   & \multicolumn{2}{c|}{NS5-NS5-F1(1)} & \multicolumn{1}{c|}{F1-D3-D5}  & \multicolumn{2}{c|}{NS5-D1-D3}  & \multicolumn{1}{c|}{M2-M5-M5(1)} \\
\multicolumn{1}{|c|}{F1-D2-D6}   & \multicolumn{2}{c|}{NS5-D2-D4(1)} & \multicolumn{1}{c|}{F1-D1-D7}  & \multicolumn{2}{c|}{NS5-D3-D5(2)} & \multicolumn{1}{c|}{}\\
\multicolumn{1}{|c|}{F1-D4-D4}   & \multicolumn{2}{c|}{NS5-D4-D6(3)} & \multicolumn{1}{c|}{}      & \multicolumn{2}{c|}{NS5-D5-D7(4)} & \multicolumn{1}{c|}{}\\
\multicolumn{1}{|c|}{}       & \multicolumn{2}{c|}{NS5-D6-D8(5)} & \multicolumn{1}{c|}{}      &         &         & \multicolumn{1}{c|}{}\\ 
\hline
\end{tabular}
\caption{Possible 3-branes configurations preserving 1/4 supersymmetry}
\label{tab:1/4-susy}
\end{table}

In order to preserve $\tfrac{1}{4}$ of the supersymmetry, we will restrict our attention to the three-brane intersections in type II theories and 11d supergravity summarised in Table \ref{tab:1/4-susy}. This list is generated by requiring that the supersymmetry-preserving projections\footnote{The projections for different branes in type II theory are listed in Appendix \ref{app:susy}.} associated with any two of the branes together imply the third one. In particular this implies that the combined worldvolumes of the three branes extend to cover all spacetime dimensions. The number in the parenthesis indicates the number of spatial dimensions shared by these branes. Any one of the three branes can be chosen as the fully localised brane, while the remaining two should be mutually smeared and fill out the transverse space of the localised brane. The harmonic functions of the mutually smeared branes should coincide. 

As an instructive example, consider the \underline{D3}-F1-D5 solution, where we have underlined our choice of the fully localised brane. The associated solution reads 
\begin{equation}
  \begin{aligned}
    &\dd \hat{s}^2_{10}=H_0^{-\frac12}\left[-h^{-1}\dd t^2+h(\dd x_1^2 + \dd x_2^2 + \dd x_3^2)\right]+H_0^{\frac12}(\dd y_1^2+\cdots+\dd y_6^2)\,,\\
    &\tilde{F}_{(5)}=h \, H_0^{-2}\partial_IH_0\,\dd t\wedge \dd x_1 \wedge \dd x_2 \wedge \dd x_3 \wedge \dd y^I - \frac{1}{5!}\varepsilon_{I_1\cdots I_6}\partial_{I_1} H_0\,\dd y^{I_2}\cdots\wedge\dd y^{I_6}\,,\\
    &\hat{H}_{(3)}=\partial_a h^{-1}\, \dd t\wedge\dd x^a \wedge\dd y^1\,,\qquad \tilde{F}_{(3)}=-\frac{1}{2!}\varepsilon_{abc}\partial_a h \,\dd x^b\wedge \dd x^c\wedge \dd y^1\,,\\
    &e^{\hat{\Phi}}=h^{-1}\,,\qquad h=h(x_1,x_2,x_3)\,,\qquad H_0=H_0(y_1,\cdots,y_6)\,.
  \end{aligned}
  \label{eq:D3-F1-D5}
\end{equation}
Here, $H_0$ is the harmonic function of the localised D3 brane and $h$ is the harmonic function of both the F1-string and the D5 brane. These satisfy 
\begin{equation}
  \partial_I^2H_0(y)=0\,,\qquad \partial_a^2 h(x)=0\,.
\end{equation}
It should be noted that the 5-form flux is not closed but is self-dual and is related to the form-field potential $\hat{C}_{(4)}$ via the covariant expression
\begin{equation}
  \tilde{F}_{(5)}=\dd \hat{C}_{(4)}-\frac{1}{2}\hat{C}_{(2)}\wedge \hat{H}_{(3)}+\frac{1}{2}\hat{B}_{(2)}\wedge \dd\hat{C}_{(2)}\,.
\end{equation}

In this system, we have a fully localised \underline{D3}-brane and a mutually smeared F1-D5 brane-pair that fills out the transverse space of the fully localised \underline{D3}-brane. This configuration is shown in Table \ref{tab:D3-F1-D5}, where \checkmark\ labels the worldvolume direction, $\sim$ indicates smearing directions and $\bullet$ denotes the localised directions where the unsmeared coordinate dependence is found. To make this interpretation of the background \eqref{eq:D3-F1-D5} manifest we may rewrite the metric as 
\begin{equation}
  \begin{aligned}
    \dd \hat{s}^2_{10} = -H_0^{-
    \frac12}H_1^{-\frac34}H_2^{-\frac14}\dd t^2 + H_0^{-
    \frac12}H_1^{\frac14}H_2^{\frac34}(\dd x_1^2 + \dd x_2^2 + \dd x_3^2) \\
    +H_0^{\frac12}H_1^{-\frac34}H_2^{\frac34}\dd y_1^2 + H_0^{\frac12}H_1^{\frac14}H_2^{-\frac14}(\dd y_2^2 + \cdots + \dd y_6^2)\,,
  \end{aligned}
  \label{eq:D3-F1-D5-metric}
\end{equation}
with $H_1=H_2=h$ being the harmonic function of the F1-string and the D5-brane, respectively. One can see that the metric \eqref{eq:D3-F1-D5-metric} and the dilaton in \eqref{eq:D3-F1-D5} indeed obey the harmonic function rule aside from the coordinate dependence of the harmonic functions at play. When we turn towards the form-fields, however, we observe the appearance of a harmonic function $h$ dressing the electric component of $\tilde{F}_{(5)}$-flux in \eqref{eq:D3-F1-D5}, which is not expected from a na\"{\i}ve application of the usual harmonic function rule. We therefore need to generalise the harmonic function rule in this context:

\bigskip
\setlength{\fboxsep}{10pt} 
\noindent\fbox{
 \parbox{0.95\linewidth}{
  \begin{enumerate}[leftmargin=1.5em]
   \item \textbf{Metric}: Write the metric following the standard harmonic function rule. The two smeared branes should share a common harmonic function $h$ that depends on the worldvolume coordinates of the fully localised brane. The harmonic function $H_0$ of the localised brane solves the flat-space Laplace equation in its transverse space.
   \item \textbf{Dilaton}: Multiply the exponential contributions to the dilaton from the different branes, as in the standard harmonic function rule.
   \item \textbf{Fluxes}: Construct the fluxes following the standard harmonic function rule. When more than two types of flux are turned on, the highest-rank RR flux (na\"{\i}vely proportional to the derivative of $h$ or $H_0$), must be multiplied by the opposite harmonic function ($H_0$ or $h$, respectively). 
  \end{enumerate}
 }
}
\bigskip

The flux corrections have been found to work empirically by considering all possible cases of brane intersections and adjusting the ansatz to generate a full solution to the 10d/11d theories. In most cases, they are due to multi-flux terms in the equations of motion and Bianchi identities. For example, the Bianchi identity of $\tilde{F}_{(4)}$ in IIA supergravity is 
\begin{equation}
  \dd \tilde{F}_{(4)} = \hat{F}_{(2)}\wedge \hat{H}_{(3)}\,.
\end{equation}
When all three fluxes are turned on, the $\hat{F}_{(2)}\wedge \hat{H}_{(3)}$ term contributes non-trivially to the 4-form flux.

In the special case of D3-branes, which are dyonic by nature, either the electric or magnetic flux is affected by harmonic factor insertions, depending on the relevant background. The electric part of the $\tilde{F}_{(5)}$-flux is corrected in the F1-D3-D5 and NS5-D1-D3 configurations (\cf \eqref{eq:D3-F1-D5}), while the magnetic part is corrected in the NS5-D3-D5(2) solutions. In these cases, the corrections to the flux are determined by requiring the $\tilde{F}_{(5)}$ self-duality constraint.

A key aspect of the new intersecting-brane solutions conforming to the above rules is the freedom of choice of the fully localised component. We have indicated the fully localised component by underlining. Instead of having the D3-brane fully localised in the \underline{D3}-F1-D5 intersecting-brane system above, one can alternatively take either of the other components to be fully localised. For example, instead of the solution \eqref{eq:D3-F1-D5} one can have a \underline{D5}-F1-D3 solution:
\begin{equation}
  \begin{aligned}
    &\dd \hat{s}^2_{10}=H_0^{-\frac14}\left[-h^{-\frac54}\dd t^2+h^{\frac34}(\dd x_1^2 + \cdots+\dd x_5^2)\right]+H_0^{\frac34}h^{-\frac14}(\dd y_1^2+\cdots+\dd y_4^2)\,,\\
    &\tilde{F}_{(5)}=H_0\, h^{-2}\partial_a h\,\dd t\wedge \dd y_2\wedge\dd y_3 \wedge \dd y_4\wedge \dd x^a - \frac{1}{4!}\varepsilon_{abcde}\partial_a h\,\dd y^1\wedge\dd x^b\wedge \dd x^c\wedge \dd x^d\wedge \dd x^e\,,\\
    &\hat{H}_{(3)}=h^{-2}\partial_a h\,\dd t\wedge\dd y^1\wedge\dd x^a\,,\qquad \tilde{F}_{(3)}=\frac{1}{3!}\varepsilon_{IJKL}\partial_I H_0\,\dd y^J\wedge \dd y^K\wedge \dd y^L\,,\\
    &e^{\hat{\Phi}}=h^{-\frac12}H_0^{-\frac12}\,,\qquad h=h(x_1,\cdots,x_5)\,,\qquad H_0=H_0(y_1,\cdots,y_4)\,.\\
  \end{aligned}
  \label{eq:D5-F1-D3}
\end{equation}
The difference between the functional dependences of the \underline{D3}-F1-D5 and the \underline{D5}-F1-D3 is illustrated in the contrast between Table \ref{tab:D3-F1-D5} and Table \ref{tab:D5-F1-D3}:
  
\begin{table}[ht]
\centering
\begin{minipage}{0.48\textwidth}
 \centering
 \begin{tabular}{|c|c|c|c|c|}
  \hline
         & t & $x_1$ - $x_3$ & $y_1$ & $y_2$ - $y_6$ \\ \hline
  \underline{D3} & \checkmark & \checkmark & $\bullet$ & $\bullet$ \\ \hline
  F1      & \checkmark & $\bullet$ & \checkmark & $\sim$ \\ \hline
  D5      & \checkmark & $\bullet$ & $\sim$ & \checkmark \\ \hline
 \end{tabular}
 \caption{\underline{D3}-F1-D5 configuration.}
 \label{tab:D3-F1-D5}
\end{minipage}
\hfill
\begin{minipage}{0.48\textwidth}
 \centering
 \begin{tabular}{|c|c|c|c|c|}
  \hline
         & t & $x_1$ - $x_5$ & $y_1$ & $y_2$ - $y_4$ \\ \hline
  \underline{D5} & \checkmark & \checkmark & $\bullet$ & $\bullet$ \\ \hline
  F1      & \checkmark & $\bullet$ & \checkmark & $\sim$ \\ \hline
  D3      & \checkmark & $\bullet$ & $\sim$ & \checkmark \\ \hline
 \end{tabular}
 \caption{\underline{D5}-F1-D3 configuration.}
 \label{tab:D5-F1-D3}
\end{minipage}
\end{table}

Following the generalised harmonic function rule, we can now construct all the solutions listed in Table \ref{tab:1/4-susy}, with all possible choices of underlinings. To present a few more examples, we consider the \underline{M5}-M2-M5(1) solution\footnote{This solution is quite similar to the M5-M2-M5 brane solution given in \cite{Tseytlin:1997cs,Gauntlett:1997cv}, which also preserves $\frac14$-supersymmetry as discussed in (\ref{eq:M5-brane-Projection}-\ref{eq:M2-brane-Projection}). However, that scenario introduces three different harmonic functions for each brane and no brane is fully localised.} in 11d supergravity:
\begin{equation}
  \begin{aligned}
    &\dd \hat{s}^2_{11}=H_0^{-\frac13}\left[h^{-1}\left(-\dd t^2+\dd x^2\right)+h(\dd r^2+r^2\dd\Omega_3^2)\right]+H_0^{\frac23}(\dd y_1^2+\cdots+\dd y_5^2)\,,\\
    &\hat{F}_{(4)}=-\frac{1}{4!}\varepsilon_{I_1\cdots I_5}\partial_{I_1}H_0\,\dd y^{I_2}\wedge\cdots\wedge\dd y^{I_5}-h^{-2}\partial_r h\,\dd t\wedge\dd t\wedge \dd r\wedge\dd y^1 - r^3 \partial_r h\,\dd \Omega_3\wedge\dd y^1\,,\\
    &h=h(r)\,,\qquad H_0=H_0(y_1,\cdots,y_5)\,,
  \end{aligned}
  \label{eq:M5-M2-M5}
\end{equation}
the \underline{D4}-F1-D4 solution in type IIA supergravity: 
\begin{equation}
  \begin{aligned}
    &\dd \hat{s}^2_{10}=H_0^{-\frac38}\left[-h^{-\frac98}\dd t^2+h^{\frac78}(\dd r^2+r^2\dd\Omega_3^2)\right]+H_0^{\frac58}h^{-\frac18}(\dd y_1^2+\cdots+\dd y_5^2)\,,\\
    &\tilde{F}_{(4)}=r^3\partial_r h\,\dd y_1\wedge \dd \Omega_3-\frac{1}{4!}\varepsilon_{I_1\cdots I_5}\partial_{I_1} H_0\,\dd y^{I_2}\wedge \cdots \dd y^{I_5}\,, \\
    &\hat{H}_{(3)}= h^{-2}\partial_r h\,\dd t\wedge \dd y_1\wedge \dd r\,,\\
    & e^{\hat{\Phi}}=h^{-\frac34}H_0^{-\frac14}\,,\qquad h=h(r)\,,\qquad H_0=H_0(y_1,\cdots,y_5)\,,
  \end{aligned}
  \label{eq:D4-F1-D4}
\end{equation}
and the \underline{NS5}-D1-D3 solution in type IIB supergravity:
\begin{equation}
  \begin{aligned}
    &\dd \hat{s}^2_{10}=H_0^{-\frac14}\left[-h^{-\frac54}\dd t^2+h^{\frac34}(\dd r^2+r^2\dd\Omega_4^2)\right]+H_0^{\frac34}h^{-\frac14}(\dd y_1^2+\cdots+\dd y_4^2)\,,\\
    &\tilde{F}_{(5)}=H_0\, h^{-2}\partial_rh\,\dd t\wedge \dd y_2\wedge\dd y_3 \wedge \dd y_4\wedge \dd r - r^4 \partial_r h\,\dd y_1\wedge\dd\Omega_4\,,\\
    &\hat{H}_{(3)}=-\frac{1}{3!}\varepsilon_{IJKL}\partial_I H_0\,\dd y^J\wedge \dd y^K\wedge \dd y^L\,,\qquad \tilde{F}_{(3)}=h^{-2}\partial_rh\,\dd t\wedge\dd y_1\wedge\dd r\,,\\
    &e^{\hat{\Phi}}=h^{\frac12}H_0^{\frac12}\,,\qquad h=h(r)\,,\qquad H_0=H_0(y_1,\cdots,y_4)\,.
  \end{aligned}
  \label{eq:NS5-D1-D3}
\end{equation}

Again, $H_0$ is the harmonic function of the fully localised brane defined with respect to the flat Laplacian in its transverse space. $h$ is the harmonic function of the two mutually-smeared branes with dependence on the worldvolume of the localised brane. In the \underline{D4}-F1-D4 case, we only have two types of fluxes, $\tilde{F}_{(4)}$ and $\hat{H}_{(3)}$, while in the \underline{M5}-M2-M5 case, we have only one type of flux $\hat{F}_{(4)}$. Hence, there are no harmonic function corrections in the fluxes in this case. In the \underline{NS5}-D1-D3 case, on the other hand, we need to correct $\tilde{F}_{(5)}$ due to the appearance of three different fluxes $\tilde{F}_{(5)},\, \hat{H}_{(3)}$ and $\tilde{F}_{(3)}$.

\section{Interpretation and embedding ansatz inferral from a worldvolume perspective}\label{Skeleton}

The original motivation for investigating such brane intersections was an interest in the worldvolume theory of the fully localised brane. In \cite{Brecher:1999xf, Erickson:2021psj} it was argued that instead of a flat worldvolume metric such as in the single-brane solution \eqref{eq:p-brane solution}, one could instead consider more general Ricci-flat geometries on the worldsheet
\begin{equation}
  \dd \hat{s}^2 = H(y)^{\frac{p-7}{8}}g_{\mu\nu}\,\dd x^\mu \dd x^\nu + H(y)^{\frac{p+7}{8}}\delta_{IJ}\,\dd y^I\dd y^J\,,\qquad \mathcal{R}_{\mu\nu}(g)=0\,.
  \label{eq:Ricci-flat-p-brane solution}
\end{equation}
Extending to greater generality, one can identify an appropriate $(p+1)$-dimensional half-maximal supergravity on the worldvolume \cite{Leung:2022nhy}, determine the uplift of its fields to 10d field configurations and show that any solution of the $(p+1)$-dimensional supergravity fully satisfies the 10d equations of motion. We then call the $(p+1)$-dimensional supergravity theory a ``consistent embedding" on the worldvolume of the brane. (The ansatz is the same as for a consistent truncation, but in this context we are not restricting attention just to the lower-dimension worldvolume theory.) The brane itself can be called a ``skeleton", suggesting that the flesh of a resulting combined solution is actually the supergravity solution sitting on top of the skeleton brane. Although in principle one may also try to construct consistent embeddings of supergravity theories with certain matter multiplets, the proposal of Ref.\,\cite{Leung:2022nhy} was that at least pure supergravity with no additional multiplets will always be consistent. Explicit examples of such consistent embeddings were given in \cite{Leung:2022nhy} and we will extend the catalogue of such examples in Section \ref{Embedding}, making use of the structures unveiled in this paper.

Although the full solution space depends on the worldvolume dimension and on the precise supergravity at hand, one generically expects $\tfrac{1}{2}$-BPS solutions such as black holes, black strings and black branes to appear as possible solutions on the skeleton worldvolume. In particular, we may consider simple static solutions of the form
\begin{equation}
  ds_{p+1}^2=h(r)^{\alpha}(-\dd t^2 + \dd x_1^2+\dots+\dd x_s^2)+h(r)^{\beta}(\dd r^2+r^2\dd\Omega_{p-s-1}^2)\,,
\end{equation}
which need to be supported by further fields from the supergravity multiplet, capturing the charge of the $\tfrac{1}{2}$-BPS object. In order to recognise such solutions as uplifts back to 10d, we require overall a 10d $\tfrac{1}{4}$-BPS solution (half of the supersymmetry is broken by the skeleton brane and another half by the black object) with two harmonic functions $H(y)$ and $h(r)$, where $h$ depends only on the world-volume coordinates of the skeleton brane, which itself is fully localised. 

It is not hard to see that the brane intersections discussed in Section \ref{Intersecting brane} are 
natural candidates for uplifts of such consistently embedded black objects on lower-dimensional worldvolumes. In fact, we claim that every such intersection may be interpreted as a solution of the worldvolume supergravity on a fully localised (skeleton) brane, featuring a $\tfrac{1}{2}$-BPS black object with a spatial extension according to the number of shared dimensions (as given in parentheses in Table \ref{tab:1/4-susy}). 

Now consider the inverse question: Can one use intersecting-brane higher-dimensional solutions as discussed in Section \ref{Intersecting brane} to infer the structure of consistent embedding ans\"atze onto the worldvolume of a fully localised component brane, now considered as a skeleton? Indeed we find that the structure of such intersecting solutions is highly restrictive and therefore implies the structure of ans\"atze for consistent embeddings of  pure supergravity on the skeleton branes.

\subsection{Pure 4d $\mathcal{N}=4$ supergravity on the D3-brane}\label{subsec:D3}

Let us demonstrate this line of thought with the now familiar \underline{D3}-F1-D5 example. We can identify the additional geometry on the skeleton D3-brane worldvolume in \eqref{eq:D3-F1-D5} with a four-dimensional black hole.

Instead of the static $\mathbb{R}^{1,3}$-worldvolume of the D3-brane, in Ref.\ \cite{Leung:2022nhy} it was found that pure $\mathcal{N}=4$ supergravity can be embedded on the 4d worldvolume, thus allowing for a curved metric $g_{\mu\nu}$, 6 $U(1)$ gauge fields $\mathcal{A}_{(1)}^i$ with 2-form fluxes $\mathcal{F}_{(2)}^i$ and a complex scalar $\tau=\chi +ie^{-\phi}$ available to be turned on. The embedding in type IIB supergravity was given as 
\begin{equation}\begin{split}\label{general}
  &\dd \hat{s} ^2_{10}= H_0(y)^{-\frac{1}{2}} \,g_{\mu\nu}\dd x^\mu \dd x^\nu + H_0(y)^{\frac{1}{2}}\,\delta_{IJ}\dd y^I \dd y^J\,,\qquad\hat{\Phi}=\phi\,,\qquad \hat{C_0}=-\chi\,,\\
  &\tilde{F}_{(5)}=-\text{vol}_{g}\wedge\dd H_0^{-1}-\ast_\delta \dd H_0\,,\quad\
  \tilde F_{(3)}=-\frac{e^{-\phi}}{\sqrt{2}}\ast_g \mathcal{F}_{(2)}^I\wedge \dd y^I\,,\quad\hat H_{(3)}=\frac{1}{\sqrt{2}}\mathcal{F}_{(2)}^I\wedge \dd y^I \,.
\end{split}\end{equation} 
Here, the volume form $\text{vol}_g$ and the Hodge star $*_g$ are associated to the 4d metric $g_{\mu\nu}$, while the Hodge $*_\delta$ is defined with respect to the 6d flat metric $\delta_{IJ}$. The harmonic function $H_0(y)$ solves the Laplace equation \eqref{eq:harmonic function} in the D3-brane transverse space. For every solution of the 4d equations of motion, this ansatz immediately solves the 10d Bianchi identities and equations of motion \cite{Leung:2022nhy}, making this embedding a consistent truncation from 10d type IIB supergravity down to pure 4d $\mathcal{N}=4$ supergravity. 

Pure 4d $\mathcal{N}=4$ supergravity allows for a plethora of supersymmetric black-hole solutions (see \cite{Neugebauer:1969wr,Breitenlohner:1987dg,Sen:1992ua,Sen:1992wi,Kallosh:1992ii,Ortin:1992ur,Kallosh:1993yg,Kallosh:1994ba,Clement:1996nh,Galtsov:1998mhf,Clement:2004ii,Bellorin:2005zc,Chen:2005uw} for a non-exhaustive list of discussions). A set of particularly simple solutions has the following non-vanishing field content
\begin{equation}
  \dd s_4^2 = -h^{-1}\,\dd t^2 + h\delta_{ab}\dd X^ a\dd X^b\,, \quad \mathcal{A}^1_{(1)}=-\sqrt{2}h^{-1}\dd t\,,\quad \phi=-\ln h\,, \label{eq:4d-bh}
\end{equation}
where we have renamed $\{x^1,x^2,x^3\}$ to capital $X$ and $h$ is a harmonic function in this 3d subspace: 
\begin{equation}
\partial_X^2h(X)=0\,.
\end{equation}
We may think of a simple single-center solution $h(X)=1+\tfrac{q}{\abs{X}}$ for visualisation. We chose $\mathcal{A}^1_{(1)}$ here for simplicity, but a global $SO(6)$-rotation allows us to turn on any linear combination of the gauge fields. 
The relevant worldsheet fluxes are 
\begin{equation}
\mathcal{F}_{(2)}^1=\sqrt{2}\dd t\wedge\dd h^{-1} \,,\quad \ast\mathcal{F}_{(2)}^1=-\frac{\sqrt{2}}{h}\ast_X\dd h\,,
\end{equation}
and the 10d embedding of this solution takes the form
\begin{equation}\label{embed}\begin{split}
\dd \hat{s} ^2_{10}&= H_0^{-\frac{1}{2}}(y)\left[-h^{-1}(X)\,\dd t^2+h(X)\,\delta_{ab}\dd X^ a\dd X^b\right] + H_0^{\frac{1}{2}}(y)\,\delta_{IJ}\dd y^I \dd y^J\,,\quad \hat{\Phi}=-\ln h\,,\\
\tilde{F}_{(5)}&=-\text{vol}_{g}\wedge\dd H_0^{-1}-\ast_\delta \dd H_0\,,\quad \tilde{F}_{(3)}=-\ast_X
(\dd h)\wedge \dd y^1\,,\quad
\hat{H}_{(3)}=\dd t\wedge\dd h^{-1} \wedge \dd y^1 \,.
\end{split}\end{equation}
We immediately recognise here the \underline{D3}-F1-D5 system of Eq.\,\eqref{eq:D3-F1-D5}. The specific choice of turning on $\mathcal{A}^1$ breaks the $SO(6)$ symmetry of the transverse space and determines that the F1 worldvolume is oriented in the dimension parametrised by $y^1$. We thus find a full $SO(6)$ orbit of equivalent black-hole/intersecting-brane solutions, which is a strict subspace of the full parameter space of supersymmetric black-hole solutions \cite{Neugebauer:1969wr,Breitenlohner:1987dg,Clement:1996nh,Galtsov:1998mhf,Leung:2022nhy}. Another simple set of black-hole solutions is related to this one by S-duality and corresponds to the \underline{D3}-D1-NS5 system in Table \ref{tab:1/4-susy}.\footnote{The remaining intersection \underline{D3}-D5-NS5(2) system in Table \ref{tab:1/4-susy} generates domain walls in 4d $\mathcal{N}=4$ supergravity and is thus not relevant for the discussion of black holes.} Black holes not falling into one of these two sets uplift to 10d objects that have yet to be described in a natural 10d language such as that of brane intersections.\footnote{More specifically, the black holes investigated here furnish a submanifold of the $SO(8,2)/(SO(6,2)\cross SO(2))$ coset used in \cite{Neugebauer:1969wr,Breitenlohner:1987dg,Clement:1996nh,Galtsov:1998mhf,Leung:2022nhy} for a sigma-model description of the parameter space of static black holes. It would be interesting to find the 10d interpretation of the full coset space.}

Now consider the inverse question: starting from the \underline{D3}-F1-D5 system of Eq.\,\eqref{eq:D3-F1-D5}, can one infer the corresponding consistent reduction ansatz on a skeleton D3 brane? From \eqref{eq:4d-bh}, one immediately recognises the 3-form flux as the wedge product of the black hole flux with $\dd y^1$. At the same time, we know that there are 6 gauge fields in the supergravity multiplet of $\mathcal{N}=4$ supergravity, transforming in the vector representation of the $SO(6)_R$ symmetry. The R-symmetry corresponds to the $SO(6)$ symmetry in the transverse space of the skeleton D3-brane. We further have knowledge of an $SO(2)$ symmetry for the dilaton and axion fields in both $\mathcal{N}=4$ supergravity and in type IIB supergravity. Hence, it is natural for us to write the embedding ansatz
\begin{equation}
  \begin{aligned}
    &\dd \hat{s}^2_{10} = H_0^{-\frac12}(y)g_{\mu\nu}(x) \dd x^\mu \dd x^\nu + H_0^{\frac12}(y)\delta_{IJ}\dd y^I \dd y^J\,,\qquad \hat{\Phi}=\phi(x)\,,\qquad \hat{C}_{(0)}=\alpha\chi(x)\,,\\
    &\tilde{F}_{(5)}=H_0^{-2}\text{vol}_g\wedge\dd H_0 - *_\delta \dd H_0\,,\quad \tilde{F}_{(3)}=-\frac{e^{-\phi}}{\sqrt2}*_g\mathcal{F}^I_{(2)}\wedge\dd y^I\,,\quad\,\hat{H}_{(3)}= \frac{1}{\sqrt2}\mathcal{F}^I_{(2)}\wedge\dd y^I.
  \end{aligned}
  \label{eq:Sugra-D3-ansatz}
\end{equation}
which is exactly the embedding ansatz of $\mathcal{N}=4$ pure supergravity given in \cite{Leung:2022nhy} up to the so far unfixed parameter $\alpha$ which relates the scalar field $\chi$ to the 10d axion $ \hat{C}_{(0)}$. One can determine the coefficient $\alpha=-1$ by plugging the ansatz into the type IIB equations of motion and requiring them to reduce to the 4d equations of motion -- this is the standard requirement of a consistent embedding. Another way, motivated instead by the 10d intersecting brane families, is to find a 10d intersecting-brane solution that is charged under the axion symmetry and to make an analogous inferral. There is no such solution in the family of solutions investigated in Section \ref{Intersecting brane} but we may instead investigate an intersecting-brane solution between a D3- and a single D7-brane given by
\begin{equation}
  \begin{aligned}
    \dd \hat{s}_{10} =& H_0^{-\frac12}\left[-\dd t^2+\dd x_0^2+h(\dd x_1^2 + \dd x_2^2)\right] + H_0^{\frac12}(\dd y_1^2+\cdots+\dd y_6^2)\,,\\
    \tilde{F}_{(5)}=& h H_0^{-2}\partial_I H_0\,\dd t\wedge \dd x_0 \wedge \dd x_1 \wedge \dd x_2\wedge \dd y^I - \frac{1}{5!}\varepsilon_{I_1\cdots I_6}\partial_{I_1} H_0\,\dd y^{I_2}\wedge\cdots\wedge \dd y^{I_6}\\
    e^{\hat{\Phi}}=&h^{-1}\,,\quad \hat{F}_{(1)}=-\varepsilon_{ab}\partial^a h\,\dd x^b\,,\quad h=h(x_1\,,x_2)\,\quad H_0=H_0(y)\,, \quad a,b=1,2\,.
  \end{aligned}
\end{equation}
It is worth noting that this solution is fully localised for both branes and on the skeleton D3 worldvolume it corresponds to a black-string solution which is charged magnetically under the axion symmetry: 
\begin{equation}
  \begin{aligned}
    \dd s^2 =& -\dd t^2 + \dd x_0^2 + h(\dd x_1^2 + \dd x_2^2)\,,\qquad e^\phi = h^{-1}\,,\\ 
    F_{(1)} =& -\varepsilon_{ab} \partial^a h\,\dd x^b\,,\qquad a,b=1,2\,. 
  \end{aligned}
\end{equation}
Requiring a matching of these solutions through a consistent embedding fixes $\alpha=-1$.
\subsection{Pure 6d $\mathcal{N}=(2,0)$ supergravity on the M5-brane}\label{subsec:M5}

A second example discussed in \cite{Leung:2022nhy} was the world-volume theory on M5-branes. Pure $\mathcal{N}=(2,0)$ supergravity may be embedded on an M5-skeleton, allowing for a curved metric and five 2-form gauge fields $\mathcal{A}^I_{(2)}$ with anti-self-dual 3-form flux. The embedding into M-theory was given as 
\begin{equation}
\begin{aligned}
    &\dd \hat s ^2_{11}= H_0^{-\frac{1}{3}}(y) g_{\mu\nu}\dd x^\mu \dd x^\nu + H_0^{\frac{2}{3}}(y)\delta_{IJ}\dd y^I \dd y^J\,,\\
    &\hat F_{(4)}=\delta_{IJ} G_{(3)}^I\wedge \dd y^J-\ast_\delta \dd H_0\,,\quad G_{(3)}^I=-\ast_g G_{(3)}^I\,.
\end{aligned}
\label{eq:6d-M5-ansatz}
\end{equation} 
Here, the Hodge star operator $*_g$ is associated to the 6d metric $g_{\mu\nu}$, while the Hodge $*_\delta$ is defined according to the 4d flat metric $\delta_{IJ}$. The harmonic function $H_0(y)$ again solves the Laplace equation \eqref{eq:harmonic function} in the transverse space. For every solution of the 6d theory, this ansatz solves the 11d Bianchi identities and equations of motion \cite{Leung:2022nhy}, making it a consistent truncation of 11d M-theory down to a 6d $\mathcal{N}=(2,0)$ supergravity. 

Given that the only gauge field in this theory is a 2-form field, the BPS objects in this theory are black strings, e.g. 
\begin{equation}\label{Rahim}
\dd s_6^2 = h^{-1}\,(-\dd t^2 +\dd x_1^2)+ h\,\delta_{ab}\dd X^a\dd X^b\,, \quad G^1_{(3)}=-\frac{1}{h^2}\dd t\wedge\dd x^1\wedge \dd h-\ast_X\dd h\,,
\end{equation}
where we have again relabelled the coordinates $\{x^2,x^3,x^4,x^5\}$ by capital $X$.
As usual, $h$ is a harmonic function in the $X$-subspace and we may think of $h(X)=1+\tfrac{q}{\abs{X}^2}$ as the simplest non-trivial example. The choice of $G^1_{(3)}$ as the only non-trivial gauge flux is arbitrary and we may perform an $SO(5)$-rotation to mix different gauge fluxes. The 11d uplift now takes the form
\begin{equation}\label{solution}\begin{split}
\dd \hat s_{11}^2 &= H_0^{-\frac{1}{3}}(y)\left[h^{-1}(X)\,(-\dd t^2+\dd x_1^2)+h(X)\,\delta_{ab}\dd X^a\dd X^b\right] + H_0^{\frac{2}{3}}(y)\delta_{IJ}\dd y^I \dd y^J\\
\tilde F_{(4)} &= -\left[\frac{1}{h^2}\dd t\wedge\dd x^1\wedge \dd h+\ast_X\dd h\right]\wedge \dd y^1-\ast_\delta \dd H_0 \ .
\end{split}
\end{equation}
We recognise here the \underline{M5}-M2-M5(1) scenario \eqref{eq:M5-M2-M5} where the choice of the gauge field $G^1_{(3)}$ corresponds to the orientation of the M2 worldvolume. Allowing for generic orientations within the $x^\mu$ and $y^I$ subspaces, as well as generic harmonic functions $h$, it seems that in this case the correspondence of our brane intersections to $\tfrac{1}{2}$-BPS black-brane solutions in 6d $\mathcal{N}=(2,0)$ pure supergravity is one-to-one. This striking feature may be credited to the relative simplicity of M-theory. 

Once again, we can extract the consistent reduction ansatz on a skeleton M5-brane. We recognise that the fluxes supported by the harmonic function $h$ in the intersecting brane solution \eqref{eq:M5-M2-M5} can be rewritten as the wedge product of the (dual) black-string flux with $\dd y^1\,$. Recall that there are five 2-form potentials transforming in the vector representation of $SO(5)_R$. After restoring this $SO(5)_R$ symmetry by covariantly replacing the label $1$ with $I$, we recover the embedding ansatz \eqref{eq:6d-M5-ansatz}.

\section{Further consistent embeddings}\label{Embedding}
We now apply the ansatz generation strategies outlined in the previous section to construct new consistent embedding ans\"atze for the worldvolume supergravities on the D4 and the NS5 branes. This extends the survey initiated in \cite{Leung:2022nhy}. Some details on the consistent embedding of fermion fields and the full proof of consistency are delegated to Appendix \ref{app:fermion-consistency}.

\subsection{Pure 5d $\mathcal{N}=4$ supergravity on the D4-brane}
Assuming the explicit 10d intersecting brane configurations constructed above are uplifts of black-brane solutions of a lower-dimensional supergravity, we can construct the relevant embedding ansatz for that supergravity, potentially with the help of additional symmetries. Let us use this idea to construct the embedding ansatz for 5d $\mathcal{N}=4$ pure supergravity on the D4 brane in type IIA theory and 6d $\mathcal{N}=(1,1)$ supergravity on the NS5 brane in type IIB theory. We will show the bosonic ansatz in this section and provide the fermionic completion in Appendix \ref{fermion-embedding}. We explicitly prove the consistency of the truncation to 6d $\mathcal{N}=(1,1)$  supergravity in Appendix \ref{consistency}. The proof for 5d $\mathcal{N}=4$ supergravity is parallel.

According to Table \ref{tab:1/4-susy}, we can use the \underline{D4}-F1-D4, \eqref{eq:D4-F1-D4}, or \underline{D4}-NS5-D2(1) solution to construct black holes in the embedding 5d $\mathcal{N}=4$ supergravity on the D4-brane, which is described in Appendix \ref{subsec:5d-sugra}. By comparing \eqref{eq:D4-F1-D4} to the black-hole solution \eqref{eq:5d-bh} , we can immediately make an initial ansatz
\begin{equation}
  \begin{aligned}
    \dd \hat{s}_{10}^{2}&=H_{0}^{-\frac{3}{8}}(y)e^{\frac{5}{8\sqrt{6}}\sigma}g_{\mu\nu}\left(x\right)\dd x^{\mu}\dd x^{\nu}+H_{0}^{\frac{5}{8}}(y)e^{-\frac{3}{8\sqrt{6}}\sigma}\delta_{IJ}\dd y^{I}\dd y^{J}\,,\quad e^{\hat\Phi}=H_{0}^{-\frac{1}{4}}e^{-\frac{9}{4\sqrt{6}}\sigma}\,.\\
    \tilde{F}_{(4)}&=-*_{\delta}\dd H_{0}-\frac{1}{\sqrt{2}}e^{\frac{2}{\sqrt{6}}\sigma}*_{g}F_{(2)}^{1}\wedge \dd y^{1}\,,\qquad \hat{H}_{(3)}=\frac{1}{\sqrt{2}}F_{(2)}^{1}\wedge\dd y^{1}\,.
  \end{aligned}
  \label{eq:5dN4-bosonic-ansatz-1}
\end{equation}
Here $*_{\delta}$ is the Hodge dual with respect to the transverse flat space $\delta_{IJ}\dd y^{I}\dd y^{J}$, while $*_{g}$ is the Hodge dual with respect to the metric $g_{\mu\nu}$. We know there are five 1-form gauge fields in the supergravity multiplet transforming in the $\textbf{5}$ representation of $USp(4)_R$.
This R-symmetry group is locally isomorphic (\ie at the level of the Lie algebra) to the rotational symmetry group $SO(5)$ in the transverse space of the D4 brane. Hence, we should replace the label for the flux $F^1_{(2)}$ with a $SO(5)$-covariant index $I\in\{1,2,3,4,5\}$. There is another 2-form flux $G_{(2)}$ which is invariant under the $USp(4)_R$, while there is a unique 2-form flux $\hat{F}_{(2)}$ in IIA theory. It is thus natural to expect 
\begin{equation}
  \hat{F}_{(2)} = \alpha G_{(2)}\,.
  \label{eq:G2-coefficient}
\end{equation} 
After a quick check of the consistent reduction of the equations of motion, we deduce the full bosonic ansatz
\begin{equation}
  \begin{aligned}
    \dd \hat{s}_{10}^{2}&=H_{0}^{-\frac{3}{8}}(y)e^{\frac{5}{8\sqrt{6}}\sigma}g_{\mu\nu}\left(x\right)\dd x^{\mu}\dd x^{\nu}+H_{0}^{\frac{5}{8}}(y)e^{-\frac{3}{8\sqrt{6}}\sigma}\delta_{IJ}\dd y^{I}\dd y^{J}\,,\quad e^{\hat{\Phi}}=H_{0}^{-\frac{1}{4}}e^{-\frac{9}{4\sqrt{6}}\sigma}\,\\
    \tilde{F}_{(4)}&=-*_{\delta}\dd H_{0}-\frac{1}{\sqrt{2}}e^{\frac{2}{\sqrt{6}}\sigma}*_{g}F_{(2)}^{I}\wedge dy^{I}\,,\quad\hat{H}_{(3)}=\frac{1}{\sqrt{2}}F_{(2)}^{I}\wedge\dd y^{I}\,,\quad\hat{F}_{(2)}=-G_{(2)}\,.
  \end{aligned}
  \label{eq:5dN4-bosonic-ansatz}
\end{equation}
This agrees with the bosonic embedding ansatz given in \cite{Leung:2022nhy}, which was inferred via a dimensional reduction of the embedding ansatz for 6d $\mathcal{N}=(2, 0)$ supergravity on the M5 brane. We will also present the embedding of fermions to leading order in Appendix \ref{fermion-embedding}.

Another way to determine the coefficient $\alpha$ in \eqref{eq:G2-coefficient} is by using the D0-D4 brane solution
\begin{equation}
  \begin{aligned}
    &\dd s^2=-H_0^{-\frac38}[h^{-\frac78}\dd t^2 +h^{\frac18}(\dd r^2+r^2\dd\Omega_3^2)]+H_0^{\frac58}h^{\frac18}(\dd y_1^2 +\cdots+\dd y_5^2)\,,\\
    &\tilde{F}_{(4)}=-*_{\delta}\dd H_0\,,\qquad \hat{F}_{(2)}=-h^{-2}\partial_rh\,\dd t\wedge\dd r\,,\qquad e^{\hat{\Phi}}=H_0^{-\frac14}h^{\frac34}\,,\\
    &H_0=H_0(y)\,,\qquad h=h(r)\,,
  \end{aligned}
\end{equation}
which corresponds to a black-hole solution \eqref{eq:5d-bh-g} charged under $G_{(2)}$ in 5d. Or, similarly, one may use a string solution charged under $G_{(2)}$\,.

\subsection{Pure 6d $\mathcal{N}=(1,1)$ supergravity on the NS5-brane of type IIB theory}\label{NS5}
By a similar analysis comparing the \underline{NS5}-D1-D3 solution \eqref{eq:NS5-D1-D3} with the black-hole solution in 6d $\mathcal{N}=(1,1)$ supergravity \eqref{eq:6d-bh}, we can obtain an initial ansatz 
\begin{equation}
  \begin{aligned}
    \dd \hat{s}_{10}^{2}=&H_{0}^{-\frac{1}{4}}(y)e^{-\frac{\sqrt{2}}{4}\varphi}g_{\mu\nu}\left(x\right)\dd x^{\mu}\dd x^{\nu}+H_{0}^{\frac{3}{4}}(y)e^{\frac{\sqrt{2}}{4}\varphi}\delta_{IJ}\dd y^{I}\dd y^{J}\,,\\
    \hat{H}_{(3)}=&-*_{\delta}\dd H_{0}\,,\qquad\tilde{F}_{(3)}=-\frac{1}{\sqrt{2}}F_{(2)}^{1}\left(x\right)\wedge\dd y^{1}\,,\qquad e^{\hat{\Phi}}=H_{0}^{\frac{1}{2}}e^{-\frac{\varphi}{\sqrt{2}}}\,,\\
    \tilde{F}_{(5)}=&\frac{1}{\sqrt{2}}\left[-\frac{1}{3!}H_{0}\varepsilon_{1 I_{2}\cdots I_{4}}F_{(2)}^{1}\wedge\dd y^{I_{2}}\wedge\dd y^{I_{3}}\wedge \dd y^{I_{4}}+e^{-\frac{\sqrt{2}}{2}\varphi}*_{g}F_{(2)}^{1}\wedge\dd y^{1}\right]\,.\\
  \end{aligned}
\end{equation}
The subscripts $\delta$ and $g$ have the same meaning as in the 5d case \eqref{eq:5dN4-bosonic-ansatz-1}. We know there are 4 gauge fields in the supergravity multiplet transforming in the vector representation of $SO(4)_R$, which is again the rotational symmetry in the transverse space of the NS5 brane. This suggests again replacing the index $1$ with $I$. In addition, there is an $SO(4)_R$ singlet 2-form gauge field $B_{(2)}$, which can be naturally embedded as
\begin{equation}
  \hat{H}_{(3)} =\alpha H_{(3)}(y)\,.
  \label{eq:H3-coefficient}
\end{equation}
After a quick check of the consistent reduction of the equations of motion, we fix $\alpha$ and get\footnote{Actually, the sign associated to the flux $F_{(2)}^I$ in the ansatz does not matter for the equations of motion, because the $F_{(2)}^I$ always appear in pairs. However, the sign does matter in the supersymmetry transformation and is related to the sign of the intertwiner.} 
\begin{equation}
  \begin{aligned}
    \dd \hat{s}_{10}^{2}=&H_{0}^{-\frac{1}{4}}(y)e^{-\frac{\sqrt{2}}{4}\varphi}g_{\mu\nu}\left(x\right)\dd x^{\mu}\dd x^{\nu}+H_{0}^{\frac{3}{4}}(y)e^{\frac{\sqrt{2}}{4}\varphi}\delta_{IJ}\dd y^{I}\dd y^{J}\,,\\
    \hat{H}_{(3)}=&-*_{\delta}\dd H_{0}- H_{(3)}\left(x\right)\,,\quad\tilde{F}_{(3)}=-\frac{1}{\sqrt{2}}F_{(2)}^{I}\left(x\right)\wedge\dd y^{I}\,,\quad e^{\hat{\Phi}}=H_{0}^{\frac{1}{2}}e^{-\frac{\varphi}{\sqrt{2}}}\,,\\
    \tilde{F}_{(5)}=&\frac{1}{\sqrt{2}}\left[-\frac{1}{3!}H_{0}\varepsilon_{I_{1}\cdots I_{4}}F_{(2)}^{I_{1}}\wedge\dd y^{I_{2}}\wedge\dd y^{I_{3}}\wedge\dd y^{I_{4}}+e^{-\frac{\sqrt{2}}{2}\varphi}*_{g}F_{(2)}^{I}\wedge\dd y^{I}\right]\,.
  \end{aligned}
  \label{eq:6d11-bosonic-ansatz}
\end{equation}
The fermionic embedding ansatz and the consistency proof are given in Appendix \ref{app:fermion-consistency}. 

Another way to determine the coefficient $\alpha$ in \eqref{eq:H3-coefficient} is by comparing the F1-NS5 solution
\begin{equation}
  \begin{aligned}
    &\dd \hat{s}_{10}^{2}=H_{0}^{-\frac{1}{4}}h^{-\frac14}[h^{-\frac12}\left(-\dd t^2 + \dd x^2\right)+h^{\frac12}\left(\dd r^2 + r^2 \dd \Omega_{3}^2\right)]+H_{0}^{\frac{3}{4}}h^{\frac14}\delta_{IJ}\dd y^{I}\dd y^{J}\,,\\
    &\hat{H}_{(3)}=-*_{\delta}\dd H_{0}- h^{-2}\partial_{r}h\,\dd t\wedge\dd x\wedge\dd r\,,\qquad e^{\hat{\Phi}}=H_{0}^{\frac{1}{2}}h^{-\frac12}(r)\,,\\
    &H_0=H_0(y)\,,\qquad h=h(r)\,,\qquad I,J=1,2,3,4\,,
  \end{aligned}
  \label{eq:6d11-bosonic-ansatz-F1}
\end{equation}
with the string solution \eqref{eq:6d-string}, electrically charged under $H_{(3)}$ in 6d.

While studying these examples, we recognise a general strategy: \emph{We can use the intersecting-brane solutions given in Table \ref{tab:1/4-susy} to construct consistent truncations of type II supergravities to various supergravities on skeleton branes.}

\

Of course there are further relationships among the various brane solutions which we may make use of.
For example, we can use S-duality to transform 6d $\mathcal{N}=(1, 1)$ supergravity on an NS5-brane to 6d $\mathcal{N}=(1, 1)$ supergravity on the D5-brane
\begin{equation}
  \begin{aligned}
    \dd \hat{s}_{10}^{2}=&H_{0}^{-\frac{1}{4}}(y)e^{-\frac{\sqrt{2}}{4}\varphi}g_{\mu\nu}\left(x\right)\dd x^{\mu}\dd x^{\nu}+H_{0}^{\frac{3}{4}}(y)e^{\frac{\sqrt{2}}{4}\varphi}\delta_{IJ}\dd y^{I}\dd y^{J}\,,\\
    \hat{H}_{(3)}=&\frac{1}{\sqrt{2}}F_{(2)}^{I}\left(x\right)\wedge\dd y^{I}\,,\quad\tilde{F}_{(3)}=-*_{\delta}\dd H_{0}- H_{(3)}\left(x\right)\,,\quad e^{\hat{\Phi}}=H_{0}^{-\frac{1}{2}}e^{\frac{\varphi}{\sqrt{2}}}\,,\\
    \tilde{F}_{(5)}=&\frac{1}{\sqrt{2}}\left[-\frac{1}{3!}H_{0}\varepsilon_{I_{1}\cdots M_{I}}F_{(2)}^{I_{1}}\wedge\dd y^{I_{2}}\wedge\dd y^{I_{3}}\wedge\dd y^{I_{4}}+e^{-\frac{\sqrt{2}}{2}\varphi}*_{g}F_{(2)}^{I}\wedge\dd y^{I}\right]\,.
  \end{aligned}
  \label{eq:6d11-bosonic-ansatz-D5}
\end{equation}
One can obtain the same reduction ansatz via an analogous construction starting from the \underline{D5}-F1-D3 brane solution \eqref{eq:D5-F1-D3}. An interesting result arises when we consider T-duality on a worldvolume direction of the NS5-brane. We know that T-duality relates NS5-branes in two type II theories. At the same time, all the configurations including an NS5-brane in Table \ref{tab:1/4-susy} are related through T-duality. This implies that pure 6d $\mathcal{N}=(1, 1)$ supergravity on an NS5-brane in type IIB theory and some version of 6d $\mathcal{N}=(2,0)$ supergravity on an NS5-brane in type IIA theory are related by T-duality. However, the number of degrees of freedom is 32 + 32 in the supergravity multiplet of $\mathcal{N}=(1, 1)$ theory while the $\mathcal{N}= (2,0)$ supergravity multiplet contains only 24 + 24 degrees of freedom. This suggests that the T-dual 6d $\mathcal{N}=(2,0)$ supergravity contains an additional tensor multiplet, which has 8 + 8 numbers of degrees of freedom.
Additional evidence for the appearance of this matter multiplet is provided by the \underline{NS5}-D4-D6(3) configuration in Table \ref{tab:1/4-susy}. The 3-brane solution in 6d should couple magnetically to a scalar which lives in a tensor multiplet. Hence, we conjecture that one can consistently embed at least one additional tensor multiplet besides the supergravity multiplet on the NS5-brane in type IIA supergravity. This should be contrasted to the pure 6d $\mathcal{N}=(2,0)$ supergravity on an NS5-brane that was given in \cite{Leung:2022nhy} through dimension reduction. We may interpret the latter as a (secondary) consistent truncation of the former.

Investigating the inclusion of matter multiplets in addition to the supergravity multiplet in the consistent truncation will be explored in future work. 

\section{Conclusion and Outlook}\label{outlook}
We have seen in this paper how the existence of consistent truncation/embedding ans\"atze from a higher-dimensional host supergravity down to lower-dimensional supergravities on ``skeleton'' brane worldvolumes allows for the construction of a new family of intersecting-brane solutions of the host theory, and, conversely, how such new intersecting-brane solutions allow one to infer the details of embedding ans\"atze. 

We started from the perspective of Ref.\,\cite{Leung:2022nhy}, which focussed on the embedding of a pure worldvolume supergravity governed by the unbroken supersymmetry of the skeleton brane. By considering static supersymmetric black-brane solutions on the worldvolume, we found families of intersecting branes with functional dependences differing from the original intersecting-brane constructions of Refs \cite{Tseytlin:1996bh,Gauntlett:1996pb}. In contrast to such constructions based on a ``harmonic function rule'' and mutually smeared branes, the new brane intersections feature the full localisation (\ie full unsmeared transverse space dependence) of one special brane. The new family of solutions agrees in generic metric structure with the original harmonic function rule, but with differing transverse functional dependences and differing ansatz details for the related form-field field strengths, as explained in Section \ref{Intersecting brane}. In Table \ref{tab:1/4-susy} we listed the possible brane intersections to preserve $\tfrac{1}{4}$ supersymmetry. We discussed also the possibility of choosing any one of the three components in these intersections to be the fully localised one, indicated here by a choice of underlining such as the distinction between the \underline{D5}-F1-D3 and the \underline{D3}-F1-D5 solutions.

Although this family of solutions already captures plenty of black objects in the worldvolume theories of the respective fully localised branes, it generically probes only a small subset of all possible $\tfrac{1}{2}$-BPS solutions in the lower-dimensional theories. This is quite obvious in the case of the D3-brane worldvolume discussed in Subsection \ref{subsec:D3}, where even the set of static black-hole solutions features many more cases \cite{Neugebauer:1969wr,Breitenlohner:1987dg,Clement:1996nh,Galtsov:1998mhf,Leung:2022nhy}. If we further allow for rotating black holes, a large landscape of solutions opens up, which can be lifted to 10d and still awaits a full interpretation in terms of 10d objects such as brane intersections. It would be an interesting challenge to further explore the higher-dimensional interpretations of black objects in various-dimensional supergravity theories. So far, only the discussion of black strings in the 6d theory on an M5-brane seems completed (see Subsection \ref{subsec:M5}).

Having understood the higher-dimensional nature of such black objects in lower-dimensional supergravity theories as brane intersections, a natural question concerns the limit of large numbers of stacked branes, in particular in the near-horizon limit of the black object. This is similar in spirit to taking the near-horizon limit of a D1-D5 brane intersection that leads to the famous $AdS_3\cross S^3\cross \mathbb{T}^4$ background after compactification. The AdS/CFT correspondence \cite{Maldacena:1997re,Witten:1998qj} may then be applied to relate the worldvolume supergravity theory to a conformal field theory on the boundary of the AdS subspace that arises in the near-horizon limit. One may therefore suspect that our new brane intersections should allow for a holographic description, but the dual theories remain to be investigated.

\vspace{10pt}

We further elaborated that, given the multitude of new intersecting-brane solutions, the structure of the related consistent truncation/embedding ans\"atze could conversely be inferred. This provides an important guide to the construction of a variety of such ans\"atze, whose full consistency can then be checked between the lower-dimensional wordvolume supergravity equations of motion and the original higher-dimensional host equations of motion. We made use of this data by consistently embedding pure 5d $\mathcal{N}=4$ supergravity on the D4-brane and pure 6d $\mathcal{N}=(1,1)$ supergravity on the NS5 brane in type IIB theory. Further such embeddings can be investigated in a similar manner.

Of particular interest are solutions with full localisation of the lowest-dimensional component such as \underline{F1}-D5-D3 and its relation to a low-dimensional consistent ``supergravity'' embedding. The worldvolume of an F1-brane is just two-dimensional, and the corresponding locally supersymmetric theory does not contain dynamical ``gravity''  -- it is instead an interacting theory of spinors and scalars. Such low-dimensional theories also have rich duality-symmetry structures, such as those discussed in Ref.\,\cite{Nicolai:1988jb}. Interpreting the uplift of solutions to such low-dimensional theories is another rich domain for further study.

The limitation to pure supergravity on the skeleton-brane worldvolume in this paper most likely does not capture the complete set of possibilities. An analogous question relates to dimensional reductions on Calabi-Yau manifolds, where truncation consistency is known for lower-dimensional pure supergravity \cite{Lin:2024eqq}. It is known from generalised-geometry considerations, however, that extensions of the reduction ans\"atze to include certain independent matter multiplets are also possible \cite{Cassani:2019vcl}, although the sorts of explicitly detailed embedding ans\"atze that we have used (or inferred) here have not been fully established in such extended cases. It is likely, moreover, that similar extensions to include some matter multiplets in the brane-worldvolume embedding ans\"atze will also exist. We found first hints of such extensions in the discussion of T-duality of the NS5 brane (at the end of Subsection \ref{NS5}),  where an additional tensor multiplet may appear in the dual theory. The related generalised-geometry analysis for the skeleton brane backgrounds has not yet been developed. Such analysis can also be the basis of a formal proof of ansatz consistency -- both for the purely bosonic sector and also when the fermions are included. These further developments will be the focus of future publications.

\section*{Acknowledgments}
We thank Zihan Wang for her input during the early development of this work. We are grateful to Costas Bachas, Jerome Gauntlett, Yusheng Jiao, Rahim Leung and Dan Waldram for helpful discussions. The work of KSS was supported in part by the STFC under consolidated grants ST/T000791/1 and ST/X000575/1. TS would like to thank the Cluster of Excellence EXC 2121 Quantum Universe 390833306 and the Collaborative Research
Center SFB1624 for creating a productive research environment at DESY.

\newpage

\begin{appendices}
\section{Conventions}\label{app:conventions}
In this appendix we summarise our conventions for supergravity theories in various dimensions. For clarity, we also outline some initial assumptions and choices made throughout the paper.
\begin{itemize}
  \item We work in ``mostly plus" signature $(-,+,+,+,\dots)$. 
  \item We discuss 10d supergravity in Einstein frame.
  \item We denote 10/11d quantities with hats and lower-dimensional quantities without hats when we are making a consistent truncation/embedding. 
  \item The local Lorentz frame is indicated by the underlined indices ($\underline{M}, \underline{N}, \dots$ for 10/11d, $\underline{\mu}, \underline{\nu},\dots$ for worldvolume directions, and $\underline{I}, \underline{J},\dots$ for transverse directions). 
  \item The 10/11d Clifford algebra is denoted by $\Gamma$. We use $\gamma$ on the worldvolume directions and $\Sigma$ on the transverse directions.
  \item We define the top rank gamma matrix as $\Gamma^{(d)}\equiv\Gamma_{\underline{0}}\Gamma_{\underline{1}}\cdots\Gamma_{\underline{d-1}}$.
  \item The totally antisymmetric symbol is chosen as $\varepsilon_{01\dots d} = 1$.
\end{itemize}

\subsection{10d type IIA supergravity}\label{subsec:IIA-sugra}
In 10d type IIA supergravity, the field content consists of the graviton $\hat{e}_{\mu}^{a}$, a Majorana gravitino $\hat{\Psi}_{\mu}$, a dilaton $\hat{\Phi}$, a Majorana dilatino $\hat{\lambda}$, an NS-NS two-form potential $\hat{B}_{(2)}$ with field strength $\hat{H}_{(3)} = \mathrm{d}\hat{B}_{(2)}$, and R-R potentials $\hat{C}_{(1)}$ and $\hat{C}_{(3)}$, with field strengths given by $\hat{F}_{(n+1)} = \mathrm{d}\hat{C}_{(n)}$. The bosonic type IIA action in Einstein frame is \cite{Campbell:1984zc}
\begin{equation}
  \begin{aligned}
    S_{IIA}=& \frac{1}{2\kappa^2}\int\left(*\hat{R} - \frac12\dd\hat{\Phi}\wedge*\dd\hat{\Phi} - \frac{1}{2}e^{-\hat{\Phi}}\hat{H}_{(3)}\wedge*\hat{H}_{(3)} - \frac{1}{2}e^{\frac32\hat{\Phi}}\hat{F}_{(2)}\wedge*\hat{F}_{(2)}\right. \\
    &\left.\qquad\qquad\qquad- \frac{1}{2}e^{\frac{\hat{\Phi}}{2}}\tilde{F}_{(4)}\wedge*\tilde{F}_{(4)}\right) - \frac{1}{4\kappa^2} \int \hat{F}_{(4)}\wedge\hat{F}_{(4)}\wedge\hat{B}_{(2)}\,.
  \end{aligned}
  \label{eq:IIA-lagrangian}
\end{equation}
Here, we have defined $\tilde{F}_{(4)}=\hat{F}_{(4)} + \hat{C}_{(1)}\wedge\hat{H}_{(3)}$ and use a hat to indicate 10d quantities. The corresponding equations of motion are 
\begin{equation}
  \begin{aligned}
  \dd*\dd\hat{\Phi}=& \frac{3}{4}e^{\frac{3}{2}\hat{\Phi}}\hat{F}_{(2)}\wedge*\hat{F}_{(2)}-\frac{1}{2}e^{-\hat{\Phi}}\hat{H}_{(3)}\wedge*\hat{H}_{(3)}+\frac{1}{4}e^{\frac{\hat{\Phi}}{2}}\tilde{F}_{(4)}\wedge*\tilde{F}_{(4)} \,,\\
   \dd\left(e^{\frac{\hat{\Phi}}{2}}*\tilde{F}_{(4)}\right)=& -\hat{H}_{(3)}\wedge\tilde{F}_{(4)}\,, \\
   \dd\left(e^{-\hat{\Phi}}*\hat{H}_{(3)}\right)=& e^{\frac{\hat{\Phi}}{2}}\hat{F}_{(2)}\wedge*\tilde{F}_{(4)}+\frac{1}{2}\tilde{F}_{(4)}\wedge\tilde{F}_{(4)}\,, \\
   \dd\left(e^{\frac{3\hat{\Phi}}{2}}*\hat{F}_{(2)}\right)=& -e^{\frac{\hat{\Phi}}{2}}\hat{H}_{(3)}\wedge*\tilde{F}_{(4)} \,,\\
  \hat{R}_{MN}=&\frac{1}{2}\partial_{M}\hat{\Phi}\partial_{N}\hat{\Phi}+\frac{1}{2}e^{\frac{3\hat{\Phi}}{2}}\left(\hat{F}_{MA}\hat{F}_{N}^{\ A}-\frac{1}{8\cdot2!}g_{MN}\hat{F}_{(2)}^{2}\right) \\
   & \qquad\quad+\frac{1}{4}e^{-\hat{\Phi}}\left(\hat{H}_{M}^{\ P_{1}P_{2}}\hat{H}_{NP_{1}P_{2}}-\frac{1}{2\cdot3!}\hat{g}_{MN}\hat{H}_{(3)}^{2}\right) \\
   & \qquad\quad+\frac{1}{12}e^{\frac{\hat{\Phi}}{2}}\left(\tilde{F}_{M}^{\ P_{1}P_{2}P_{3}}\tilde{F}_{NP_{1}P_{2}P_{3}}-\frac{9}{4\cdot4!}\hat{g}_{MN}\tilde{F}_{(4)}^{2}\right) \,.
  \end{aligned}
  \label{eq:IIA-eom}
\end{equation}
The supersymmetry transformations are
\cite{becker_becker_schwarz_2006}
\begin{equation}
  \begin{aligned}
    \delta \hat{\Psi}_M =& \left(\hat{D}_M - \frac{1}{4\cdot2!} \hat{H}_{MPQ}\Gamma^{PQ}\Gamma^{(10)} \right.\\
    &\qquad \left.- \frac{1}{16} e^{\hat{\Phi}} \hat{F}_{PQ}\Gamma^{PQ}\Gamma_M\Gamma^{(10)} + \frac{1}{8\cdot4!}e^{\hat{\Phi}} \tilde{F}_{NPQL}\Gamma^{NPQL} \Gamma_M \right)\hat{\epsilon}\,, \\
    \delta \hat{\lambda} =& \left(-\frac13\Gamma^M\partial_M\hat{\Phi}\Gamma^{(10)}+\frac{1}{6\cdot3!}\hat{H}_{MNP}\Gamma^{MNP}\right.\\
    &\qquad \left.-\frac{1}{2\cdot4}e^{\hat{\Phi}} \hat{F}_{MN}\Gamma^{MN} + \frac{1}{12\cdot4!}e^{\hat{\Phi}} \tilde{F}_{MNPQ}\Gamma^{MNPQ}\Gamma^{(10)}\right)\hat{\epsilon}\,.
    \label{eq:IIA-susy-bosonic}
  \end{aligned}
\end{equation}
Here, $\Gamma^{(10)}$ is the chiral matrix in 10 dimensions. We can separate the 10d Majorana supercharges into two Majorana-Weyl supercharges with different chiralities 
\begin{equation}
  \hat{\epsilon}=\hat{\epsilon}_L + \hat{\epsilon}_R\,,\qquad \Gamma^{(10)}\hat{\epsilon}_L=\hat{\epsilon}_L\,,\qquad \Gamma^{(10)}\hat{\epsilon}_R=-\hat{\epsilon}_R\,.
  \label{eq:IIA-spinor-separate}
\end{equation}

\subsubsection{D4-brane solution}
The D4-brane solution in type IIA supergravity reads
\begin{equation}
  \begin{aligned}
    \dd s_{10}^{2}&=H^{-\frac{3}{8}}g_{\mu\nu}\left(x\right)\dd x^{\mu}\dd x^{\nu}+H^{\frac{5}{8}}\delta_{IJ}\dd y^{I}\dd y^{J}\,,\qquad e^{\Phi}=H^{-\frac{1}{4}}\left(y\right)\,,\\
    \tilde{F}_{(4)}&=-\frac{1}{4!}\partial_{I_1}H\varepsilon_{I_1\cdots I_5}\dd y^{I_2}\wedge\cdots\wedge\dd y^{I_5}\,,\qquad I,J=1,\cdots,5\,.
  \end{aligned}
\end{equation}
It admits a Killing spinor $\epsilon$ satisfying 
\begin{equation}
  \Gamma_{\underline{01234}}\epsilon=-\Gamma^{(10)}\epsilon\,,\quad\epsilon=H^{-3/32}\epsilon_{0}\,.
  \label{eq:D4-projection}
\end{equation}
where $\epsilon_{0}$ is a constant Majorana spinor. In type IIA theory, We can split the 10d gamma matrices according to
\begin{equation}
  \Gamma_{\underline{\mu}}=\sigma_{1}\otimes\gamma_{\underline{\mu}}\otimes\mathbf{1}\,,\qquad\Gamma_{\underline{I}}=\sigma_{2}\otimes\mathbf{1}\otimes\Sigma_{\underline{I}}\,,
  \label{eq:IIA-gamma-factorise}
\end{equation}
with $\Sigma_{5}=\Sigma_{1}\cdots\Sigma_{4}\,,$ and the Pauli matrices $\sigma_{i}$. We factorise the Killing spinor as 
\begin{equation}
  \epsilon=\left(\begin{array}{c}
1\\
-i
\end{array}\right)\otimes\varepsilon^{i}\otimes\eta_{i}\,,\quad i=1,2,3,4\,,
  \label{eq:D4-spinor-factorisation}
\end{equation}
where $\eta_{i}$ is a basis of constant 5d symplectic Majorana spinors and $i$ is the $USp(4)$ index. This factorisation automatically satisfies the projection property \eqref{eq:D4-projection}, which can be written as 
\begin{equation}
  \left(\sigma_{2}\otimes\mathbf{1}\otimes\mathbf{1}\right)\epsilon=-\epsilon\,.
\end{equation} 

\subsection{10d type IIB supergravity}\label{subsec:IIB-sugra}
In 10d type IIB supergravity, the field content consists of the graviton $\hat{e}_{\mu}^{a}$, a Weyl gravitino $\hat{\Psi}_{\mu}$ with positive chirality, a dilaton $\hat{\Phi}$, a Weyl dilatino $\hat{\lambda}$ with negative chirality, an NS-NS two-form potential $\hat{B}_{(2)}$ with field strength $\hat{H}_{(3)} = \mathrm{d}\hat{B}_{(2)}$, and R-R potentials $\hat{C}_{(0)}$, $\hat{C}_{(2)}$ and $\hat{C}_{(4)}$. The type IIB theory possesses a rigid $SL(2,\mathbb{R})/SO(2)$ symmetry. The pseudo-action reads \cite{becker_becker_schwarz_2006}
\begin{equation}
  \begin{aligned}
    S_{IIB}=&\frac{1}{2\kappa_{10}^{2}}\int\left(*R-\frac{1}{2}\dd\hat{\Phi}\wedge*\dd\hat{\Phi}-\frac{1}{2}e^{2\hat{\Phi}}\hat{F}_{(1)}\wedge*\hat{F}_{(1)}-\frac{1}{2}e^{-\hat{\Phi}}\hat{H}_{(3)}\wedge*\hat{H}_{(3)}\right.\\
    &\qquad\qquad\qquad\left.-\frac{1}{2}e^{\hat{\Phi}}\tilde{F}_{(3)}\wedge*\tilde{F}_{(3)}-\frac{1}{4}\tilde{F}_{(5)}\wedge*\tilde{F}_{(5)}\right)-\frac{1}{4\kappa_{10}^{2}}\int \hat{C}_{(4)}\wedge \hat{H}_{(3)}\wedge \tilde{F}_{(3)}\,,
  \end{aligned}
  \label{eq:IIB-action}
\end{equation}
where $\hat{F}_{(1)}=\dd\hat{C}_{(0)},\ \hat{H}_{(3)}=\dd B_{(2)},\ \tilde{F}_{(3)}=\dd \hat{C}_{(2)}-\hat{C}_{(0)} \hat{H}_{(3)}\ \hbox{and}\ \tilde{F}_{(5)}=\dd \hat{C}_{(4)}-\frac{1}{2}\hat{C}_{(2)}\wedge \hat{H}_{(3)}+\frac{1}{2}B_{(2)}\wedge \dd\hat{C}_{(2)}\,$.
Varying the action, we get the bosonic equations of motion 
\begin{equation}
  \begin{aligned}
    \dd*\dd\hat{\Phi}=&e^{2\hat{\Phi}}\hat{F}_{(1)}\wedge*\hat{F}_{(1)}-\frac{1}{2}e^{-\hat{\Phi}}\hat{H}_{(3)}\wedge*\hat{H}_{(3)}+\frac{1}{2}e^{\hat{\Phi}}\tilde{F}_{(3)}\wedge*\tilde{F}_{(3)}\,,\\
    \dd\left(e^{2\hat{\Phi}}*\hat{F}_{(1)}\right)=&-e^{\hat{\Phi}}\hat{H}_{(3)}\wedge*\tilde{F}_{(3)}\,,\\
    \dd\left(e^{\hat{\Phi}}*\tilde{F}_{(3)}\right)=&\tilde{F}_{(5)}\wedge \hat{H}_{(3)}\,,\\
    \dd\left(e^{-\hat{\Phi}}*\hat{H}_{(3)}\right)=&e^{\hat{\Phi}}\hat{F}_{(1)}\wedge * \tilde{F}_{(3)}-\tilde{F}_{(5)}\wedge \tilde{F}_{(3)}\,,\\
    \dd*\tilde{F}_{(5)}=&-\tilde{F}_{(3)}\wedge \hat{H}_{(3)}\,,\\
    \hat{R}_{MN}=&\frac{1}{2}\partial_{M}\hat{\Phi}\partial_{N}\hat{\Phi}+\frac{1}{2}e^{2\hat{\Phi}}\hat{F}_{M}\hat{F}_{N}+\frac{1}{4\cdot 4!}\tilde{F}_{MPQRS}\tilde{F}_{N}{}^{PQRS}\\
    &+\frac{1}{4}e^{\hat{\Phi}}\left(\hat{F}_{MPQ}\hat{F}_{N}{}^{PQ}-\frac{1}{12}G_{MN}\hat{F}_{PQR}\hat{F}^{PQR}\right)\\
    &+\frac{1}{4}e^{-\hat{\Phi}}\left(\hat{H}_{MPQ}\hat{H}_{N}{}^{PQ}-\frac{1}{12}G_{MN}\hat{H}_{PQR}\hat{H}^{PQR}\right)\,.
  \end{aligned}
  \label{eq:IIB-bosonic-eom}
\end{equation}
The supersymmetry transformations are 
\begin{equation}
  \begin{aligned}
    \delta\hat{\lambda}=&\frac{1}{2}\left(\partial_{M}\hat{\Phi}+ie^{\hat{\Phi}}\hat{F}_{M}\right)\Gamma^{M}\hat{\epsilon}-\frac{1}{4\cdot3!}\left(e^{-\hat{\Phi}/2}\hat{H}_{MNP}+ie^{\hat{\Phi}/2}\hat{F}_{MNP}\right)\Gamma^{MNP}\hat{\epsilon}^{c}\,,\\
    \delta\hat{\Psi}_{M}=&\hat{D}_{M}\hat{\epsilon}-\frac{i}{4}e^{\hat{\Phi}}\hat{F}_{M}\hat{\epsilon}+\frac{1}{96}\left(e^{-\hat{\Phi}/2}\hat{H}_{NPQ}-ie^{\hat{\Phi}/2}\hat{F}_{NPQ}\right)\left(\Gamma_{M}^{\ NPQ}-9\delta_{M}^{N}\Gamma^{PQ}\right)\hat{\epsilon}^{c}\\
    &\qquad-\frac{i}{16\cdot5!}\hat{F}_{NPQRT}\Gamma^{NPQRT}\Gamma_{M}\hat{\epsilon}\,.
  \end{aligned}
  \label{eq:IIB-susy}
\end{equation}
The gravitino and the supersymmetry parameter have the same chirality 
\begin{equation}
  \Gamma^{(10)}\hat{\Psi}_M=\hat{\Psi}_M\,\qquad \Gamma^{(10)}\hat{\epsilon}=\hat{\epsilon}\,,
  \label{eq:IIB-gravitino-chiral}
\end{equation}
while the dilatino has the opposite chirality 
\begin{equation}
  \Gamma^{(10)}\hat{\lambda}=-\hat{\lambda}\,.
  \label{eq:IIB-dilatino-chiral}
\end{equation}
We can separate the Weyl supercharges into two Majorana-Weyl supercharges of the same chirality
\begin{equation}
  \hat{\epsilon}=\hat{\epsilon}_L + i\hat{\epsilon}_R\,,\qquad \hat{\epsilon}^c=\hat{\epsilon}_L - i\hat{\epsilon}_R\,,\qquad \Gamma^{(10)}\hat{\epsilon}_{L,R}=\hat{\epsilon}_{L,R}\,.
  \label{eq:IIB-spinor-separate}
\end{equation}

\subsubsection{NS5-brane solution}
The NS5-brane solution in type IIB supergravity reads 
\begin{equation}
  \begin{aligned}
    \dd s_{10}^{2}=&H^{-\frac{1}{4}}\eta_{\mu}\dd x^{\mu}\dd x^{\nu}+H^{\frac{3}{4}}\delta_{IJ}\dd y^{I}\dd y^{J}\,,\\
    \hat{H}_{(3)}=&-*_{\delta}\dd H\,,\qquad e^{\Phi}=H^{\frac{1}{2}}\,,\qquad I,J=1,2,3,4\,.
  \end{aligned}
\end{equation}
Here, $H$ is a harmonic function in the transverse flat space. This solution preserves 16 supercharges. The Killing spinor is 
\begin{equation}
  \epsilon=-\Gamma_{\underline{012345}}\epsilon^{c}\,,\qquad\epsilon=H^{-\frac{1}{16}}\epsilon_{0}\,.
  \label{eq:6d-NS5-project}
\end{equation}
Here, in the context of type IIB theory, we factorise the 10d gamma matrix as 
\begin{equation}
  \Gamma_{\underline{\mu}}=\gamma_{\underline{\mu}}\otimes\mathbf{1}\,,\qquad\Gamma_{\underline{m}}=\gamma^{(6)}\otimes\Sigma_{\underline{m}}\,,
  \label{eq:IIB-gamma-factorise}
\end{equation}
and rewrite the Killing spinor as
\begin{equation}
  \epsilon=\varepsilon^{A}\otimes\eta_{A}(y)+i\varepsilon^{\dot{A}}\otimes\eta_{\dot{A}}(y)\,.
  \label{eq:KS-NS5}
\end{equation}
The $\varepsilon$ are constant symplectic Majorana-Weyl spinors in 6d, while the $\eta$ are Killing spinors in the 4d transverse space. $A$ and $\dot{A}$ are the $SU(2)$ indices labelling symplectic Majorana spinors with different chiralities in 6d
\begin{equation}
  \varepsilon^A=-\gamma^{(6)}\varepsilon^A\,,\qquad\varepsilon^{\dot{A}}=\gamma^{(6)}\varepsilon^{\dot{A}}\,.
\end{equation}
On the other hand, $A$ and $\dot{A}$ label the basis of the 4d symplectic Majorana-Weyl spinors, \ie one can understand $A$ and $\dot{A}$ combined as the $Spin(4)\simeq SU(2)_L\times SU(2)_R$ spinor index in the transverse 4d subspace. 
We require
\begin{equation}
  \eta_A=-\Sigma^{(4)}\eta_A\,,\qquad\eta_{\dot{A}}=\Sigma^{(4)}\eta_{\dot{A}}
  \label{eq:5d-transverse-spinor}
\end{equation}
to ensure that the Killing spinor \eqref{eq:KS-NS5} is positive chiral. One can then see that the factorisation \eqref{eq:KS-NS5} satisfies the projection condition \eqref{eq:6d-NS5-project} automatically.

\subsection{Pure 6d $\mathcal{N}=(1,1)$ supergravity}\label{subsec:6d-sugra}
The 6d $\mathcal{N}=(1,1)$ supergravity was constructed in \cite{Giani:1984dw,Romans:1985tw,Andrianopoli:2001rs} based on the F(4) Poincar\'e superalgebra. The supergravity multiplet consists of a graviton $e_{\mu}^{\ m}$, a pair of symplectic Majorana gravitini $\psi_{\mu i}$, a two-form field $B_{(2)}$, four vector fields $A_{\mu}^{I}$, a pair of symplectic Majorana dilatini $\chi_{i}$, and one real scalar field $\varphi$. Here, $I,J=0,1,2,3$ label the vector representation $\mathbf{4}$ of the $SO(4)_R$ symmetry, while $i,j$ label the symplectic $Sp(1)\simeq SU(2)$ doublet. The ungauged pure supergravity action was given in Refs \cite{Sezgin:2023hkc, Andrianopoli:2001rs}. The bosonic action reads
\begin{equation}
  \begin{aligned}
    S=&\frac{1}{2\kappa^2_6}\int*R-\frac{1}{2}e^{-\frac{\varphi}{\sqrt{2}}}\delta_{IJ}F_{(2)}^{I}\wedge*F_{(2)}^{J}-\frac{1}{2}e^{\sqrt{2}\varphi}H_{(3)}\wedge*H_{(3)}\\
    &\qquad \qquad-\frac{1}{2}\dd\varphi\wedge*\dd\varphi+\frac{1}{2}\delta_{IJ}B_{(2)}\wedge F_{(2)}^{I}\wedge F_{(2)}^{J}\,.
  \end{aligned}
\end{equation}
The equations of motion are 
\begin{equation}
  \begin{aligned}
    \dd*\dd\varphi=&-\frac{1}{2\sqrt{2}}e^{-\frac{\varphi}{\sqrt{2}}}\delta_{IJ}F_{(2)}^{I}\wedge*F_{(2)}^{J}+\frac{\sqrt{2}}{2}e^{\sqrt{2}\varphi}H_{(3)}\wedge*H_{(3)}\,,\\\dd\left(e^{-\frac{\varphi}{\sqrt{2}}}*F_{(2)}^{I}\right)=&H_{(3)}\wedge F_{(2)}^{I}\,,\\
    \dd\left(e^{\sqrt{2}\varphi}*H_{(3)}\right)=&-\frac{1}{2}\delta_{IJ}F_{(2)}^{I}\wedge F_{(2)}^{J}\,,\\
    R_{\mu\nu}=&\frac{1}{2}\partial_{\mu}\varphi\partial_{\nu}\varphi+\frac{1}{2}e^{-\frac{\varphi}{\sqrt{2}}}\delta_{IJ}\left(F_{\mu\rho}^{I}F_{\nu}^{J\rho}-\frac{1}{8}g_{\mu\nu}F_{\rho\sigma}^{I}F^{J\rho\sigma}\right)\\&+\frac{1}{4}e^{\sqrt{2}\varphi}\left(H_{\mu\rho\sigma}H_{\nu}^{\ \rho\sigma}-\frac{1}{6}g_{\mu\nu}H_{\rho\sigma\lambda}H^{\rho\sigma\lambda}\right)\,.
  \end{aligned}
\end{equation}
For the supersymmetry transformations, instead of the indices $I,J$, we explicitly use $(0,a)$, with $a$ being a fundamental $SO(3)\subset SO(4)$ index. The supersymmetry transformations are
\begin{equation}
  \begin{aligned}
    \delta\psi_{\mu i}=&D_{\mu}\varepsilon_{i}+\frac{1}{16\sqrt{2}}e^{-\frac{\varphi}{2\sqrt{2}}}\left[F_{\rho\sigma}^{0}\epsilon_{ij}\gamma^{(6)}-iF_{\rho\sigma}^{a}\epsilon_{ik}\left(\sigma^{a}\right)_{\ j}^{k}\right]\left(\gamma_{\mu}^{\ \nu\lambda}-6\delta_{\mu}^{\nu}\gamma^{\lambda}\right)\varepsilon^{j}\\
    &+\frac{1}{48}e^{\frac{\varphi}{\sqrt{2}}}H_{\nu\lambda\rho}\gamma^{(6)}\left(\gamma_{\mu}^{\ \nu\lambda\rho}-3\delta_{\mu}^{\nu}\gamma^{\lambda\rho}\right)\varepsilon_{i}\,,\\
    \delta\chi_{i}=&-\frac{1}{4}\gamma^{\mu}\partial_{\mu}\varphi\varepsilon_{i}-\frac{1}{16}e^{-\frac{\varphi}{2\sqrt{2}}}\left[F_{\rho\sigma}^{0}\epsilon_{ij}\gamma^{(6)}+iF_{\rho\sigma}^{a}\epsilon_{ik}\left(\sigma^{a}\right)_{\ j}^{k}\right]\gamma^{\rho\sigma}\varepsilon^{j}\\
    &+\frac{\sqrt{2}}{48}e^{\frac{\varphi}{\sqrt{2}}}H_{\mu\nu\lambda}\gamma^{(6)}\gamma^{\mu\nu\lambda}\varepsilon_{i}\,.
  \end{aligned}
  \label{eq:6d-susy}
\end{equation}
The $SU(2)$ doublet indices $i,j$ are raised and lowered by the antisymmetric tensors $\varepsilon_{ij}$ and $\varepsilon^{ij}$. In view of an eventual embedding in type IIB supergravity, it will be useful to separate the symplectic Majorana spinor into two symplectic Majorana-Weyl spinors\footnote{We remind the reader that the 6d Dirac representation is given in terms of complex 8-component spinors, while the Weyl representation only requires 4 complex components. Charge conjugation in 6d squares to $-1$ so we can only consider a symplectic Majorana condition, which results in a pair of real 8-component spinors. Since the charge conjugation in 6d does not change the chirality, both conditions can be superimposed for a total of 8 real degrees of freedom, see e.g. Refs \cite{Figueroa, Coimbra:2012af} for details.} with opposite chiralities and independent symplectic structures. In particular we choose a splitting of the form\footnote{
In addition to splitting the chiralities we also introduce a factor of i in the second term. This constitutes a conventional redefinition of the symplectic structure of the second fermion and will be convenient in the 10d embedding.
}
\begin{equation}
  \psi_\mu^i= \delta_A^i\psi_\mu^A+i\delta_{\dot{A}}^i\psi_\mu^{\dot{A}}\,,\qquad \chi^i=\delta_A^i\chi^A+i\delta_{\dot{A}}^i\chi^{\dot{A}}\,,\qquad \varepsilon^i=\delta_A^i\varepsilon^A+i\delta_{\dot{A}}^i\varepsilon^{\dot{A}}\,,
  \label{eq:6d-spinors-separation}
\end{equation}
where $i,A,\dot{A} = 1,2$ are $SU(2)$ indices, respectively. Here, the individual components are constrained by additional Weyl conditions of the form
\begin{equation}
  \varepsilon^A=-\gamma^{(6)}\epsilon^A\,,\qquad\epsilon^{\dot{A}}=\gamma^{(6)}\epsilon^{\dot{A}}\,,\qquad \psi_\mu^A =-\gamma^{(6)}\psi_\mu^A\,,\qquad \psi_\mu^{\dot{A}} = \gamma^{(6)}\psi_\mu^{\dot{A}}\,,
\end{equation}
for the gravitini and the supersymmetry transformation parameters and 
\begin{equation}
  \chi^A=\gamma^{(6)}\chi^A\,,\qquad \chi^{\dot{A}}=-\gamma^{(6)}\chi^{\dot{A}}\,.
\end{equation}
for the dilatini. The different treatment of the dilatini is chosen to match the initial chirality difference in an eventual 10d embedding.

We can construct black-brane solutions in 6d following \cite{Stelle:1999ljt}. The black hole background is given by
\begin{equation}
  \begin{aligned}
    &\dd s^{2}=	-H^{-\frac{3}{2}}\dd t^{2}+H^{\frac{1}{2}}\left(\dd r^{2}+r^{2}\dd\Omega_{4}^{2}\right)\,,\\
    &F_{(2)}=	\sqrt{2}\partial_{r}H^{-1}\dd t\wedge\dd r\,,\qquad e^{\varphi}=H^{-\frac{1}{\sqrt{2}}}\,,
  \end{aligned}
  \label{eq:6d-bh}
\end{equation}
while a black 2-brane generates the background 
\begin{equation}
  \begin{aligned}
    &\dd s^{2}=	H^{-\frac{1}{2}}\left(-\dd t^{2}+\dd x_{1}^{2}+\dd x_{2}^{2}\right)+H^{\frac{3}{2}}\left(\dd r^{2}+r^{2}\dd\Omega_{2}^{2}\right)\,,\\
    &F_{(2)}=\sqrt{2}r^{2}\partial_{r}h\,\dd\Omega_{2}\,,\qquad e^{\varphi}=H^{\frac{1}{\sqrt{2}}}\,.\\
  \end{aligned}
\end{equation}
Furthermore, we can construct electric string solutions
\begin{equation}
  \begin{aligned}
    &\dd s^{2}=	H^{-\frac{1}{2}}\left(-\dd t^{2}+\dd x^{2}\right)+H^{\frac{1}{2}}\left(\dd r^{2}+r^{2}\dd\Omega_{3}^{2}\right)\,,\\
    &H_{(3)}=H^{-2}\partial_{r}h\,\dd t\wedge\dd x\wedge \dd r\,,\qquad e^{\varphi}=H^{\frac{1}{\sqrt{2}}}\,,\\
  \end{aligned}
  \label{eq:6d-string}
\end{equation}
and similarly magnetically charged ones.

\subsection{5d $\mathcal{N}=4$ pure supergravity}\label{subsec:5d-sugra}
The 5d $\mathcal{N}=4$ supergravity was constructed in \cite{Romans:1985ps, Awada:1985ep, Dall_Agata_2001, Schon:2006kz}. The $\mathcal{N}=4$ supergravity multiplet consists of a graviton $e_{\mu}^{\ m}$, four symplectic Majorana gravitini $\psi_{\mu}^{i}$, six vector fields $\left(A_{\mu}^{ij},\,a_{\mu}\right)$, four symplectic Majorana dilatini $\chi^{i}$, and one real scalar field $\sigma$. The indices $i,j=1,\cdots,4$ correspond to the fundamental representation of $USp\left(4\right)_R$. The vector field $a_{\mu}$ is $USp\left(4\right)_R$ invariant, whereas the vector fields $A_{\mu}^{ij}$ transform in the $\textbf{5}$ representation of $USp\left(4\right)_R$, \ie 
\begin{equation}
  A_{\mu}^{ij}=-A_{\mu}^{ji}\,,\qquad A_{\mu}^{ij}\Omega_{ij}=0\,,
\end{equation}
with $\Omega_{ij}$ being the symplectic metric of $USp\left(4\right)_R$. The $USp\left(4\right)_R$ indices are raised and lowered by 
\begin{equation}
  T^{i}=\Omega^{ij}T_{j}\,,\qquad T_{i}=T^{j}\Omega_{ji}\,.
\end{equation}
There is a local isomorphism $\mathfrak{so}\left(5\right)\simeq \mathfrak{usp}\left(4\right)$ . We introduce $SO\left(5\right)$ indices $I=1,2,\cdots5$, the local intertwiner $L_{I}^{\ ij}$ between these groups and its inverse $L_{ij}^{\ I}\,$. The $USp\left(4\right)$ connection is given by 
\begin{equation}
  Q^{ij}=L^{Iik}\dd L_{Ik}^{\ j}\,.
\end{equation}

The action of the bosonic sector in minimal 5d $\mathcal{N}=4$ supergravity\footnote{Here, we make a rescaling of the fields in relation to Ref.\, \cite{Dall_Agata_2001} as $A_{\mu}^{I}\rightarrow\frac{1}{\sqrt{2}}A_{\mu}^{I}\,,\ a_{\mu}\rightarrow\frac{1}{\sqrt{2}}a_{\mu}\,,\ \sigma\rightarrow\frac{1}{\sqrt{2}}\sigma\,.$ Also, we have a different convention for the sign of the Ricci tensor.} reads \cite{Dall_Agata_2001}
\begin{equation}
  \begin{aligned}
    S=&\frac{1}{2\kappa^2_5}\int*R-\frac{1}{2}e^{\frac{2}{\sqrt{6}}\sigma}a_{IJ}F^{I}_{(2)}\wedge*F^{J}_{(2)}-\frac{1}{2}e^{-\frac{4}{\sqrt{6}}\sigma}G_{(2)}\wedge*G_{(2)}\\
    &\qquad\qquad -\frac{1}{2}\dd\sigma\wedge*\dd\sigma+\frac{1}{2}C_{IJ}F^{I}_{(2)}\wedge F^{J}_{(2)}\wedge a_{(1)}\,.
  \end{aligned}
  \label{eq:5DN4-bosonic-action}
\end{equation}
The supersymmetry transformations are
\begin{equation}
  \begin{aligned}
    \delta\psi_{\mu i}=&D_{\mu}\varepsilon_{i}+\frac{i}{6\sqrt{2}}e^{\frac{1}{\sqrt{6}}\sigma}L_{Iij}F_{\rho\sigma}^{I}\left(\Gamma_{\mu}^{\ \rho\sigma}-4\delta_{\mu}^{\rho}\Gamma^{\sigma}\right)\varepsilon^{j}\\&+\frac{i}{24}e^{-\frac{2}{\sqrt{6}}\sigma}G_{\rho\sigma}\left(\Gamma_{\mu}^{\ \rho\sigma}-4\delta_{\mu}^{\rho}\Gamma^{\sigma}\right)\varepsilon_{i}\,,\\\delta\chi_{i}=&-\frac{i}{2\sqrt{2}}\Gamma^{\mu}\partial_{\mu}\sigma\varepsilon_{i}+\frac{1}{2\sqrt{6}}e^{\frac{1}{\sqrt{6}}\sigma}L_{Iij}F_{\rho\sigma}^{I}\Gamma^{\rho\sigma}\varepsilon^{j}-\frac{1}{4\sqrt{3}}e^{-\frac{2}{\sqrt{6}}\sigma}G_{\rho\sigma}\Gamma^{\rho\sigma}\varepsilon_{i}\,.
  \end{aligned}
  \label{eq:5DN4-bosonic-susy}
\end{equation}
Here, the tensors satisfy
\begin{equation}
  a_{IJ}=C_{IJ}=L_{I}^{ij}L_{Jij}\,,
\end{equation}
where $a_{IJ}$ acts as a metric on the $I,J$ indices 
\begin{equation}
  L_{I}^{\ ij}=a_{IJ}L^{Jij}\,.
\end{equation}
Actually, supersymmetry requires the symmetric tensor $C_{IJ}$ to be constant. Thus, a natural choice for the intertwiner between $USp(4)$ and $SO(5)$ is 
\begin{equation}
  \left(L_{I}\right)_{\ j}^{i}=\frac{1}{2}\left(\Sigma_{\underline{I}}\right)_{\ j}^{i}\,,
  \label{eq:USp4-SO5-intertwiner}
\end{equation}
with the gamma matrix $\Sigma_{\underline{I}}$ in the $\text{Cliff}(5)$ algebra satisfying 
\begin{equation}
  \left\{ \Sigma_{\underline{I}},\Sigma_{\underline{J}}\right\} =2\delta_{\underline{IJ}}\,.
\end{equation} 
Then, we have 
\begin{equation}
  a_{IJ}=C_{IJ}=\delta_{IJ}\,.
\end{equation}
In this representation, the equations of motion are 
\begin{equation}
  \begin{aligned}
    \dd*\dd\sigma=&\frac{1}{\sqrt{6}}e^{\frac{2}{\sqrt{6}}\sigma}\delta_{IJ}F_{(2)}^{I}\wedge*F^{J}_{(2)}-\frac{2}{\sqrt{6}}e^{-\frac{4}{\sqrt{6}}\sigma}G_{(2)}\wedge*G_{(2)}\,,\\
    \dd\left(e^{-\frac{4}{\sqrt{6}}\sigma}*G_{(2)}\right)=&\frac{1}{2}\delta_{IJ}F^{I}_{(2)}\wedge F^{J}_{(2)}\,,\\
    \dd\left(e^{\frac{2}{\sqrt{6}}\sigma}*F^{I}_{(2)}\right)=&F^{I}_{(2)}\wedge G_{(2)}\,,\\
    R_{\mu\nu}=&\frac{1}{2}\partial_{\mu}\sigma\partial_{\nu}\sigma+\frac{1}{2}\delta_{IJ}e^{\frac{2}{\sqrt{6}}\sigma}\left(F_{\mu\rho}^{I}F_{\nu}^{J\rho}-\frac{1}{6}g_{\mu\nu}F_{\rho\sigma}^{I}F^{J\rho\sigma}\right)\\&+\frac{1}{2}e^{-\frac{4}{\sqrt{6}}\sigma}\left(G_{\mu\rho}G_{\nu}^{\rho}-\frac{1}{6}g_{\mu\nu}G_{\rho\sigma}G^{\rho\sigma}\right)\,.
  \end{aligned}
  \label{eq:5DN4-bosonic-eom}
\end{equation}

This theory admits black-hole solutions \cite{Stelle:1999ljt}
\begin{equation}
  \begin{aligned}
    \dd s_{5}^{2}&=-h^{-4/3}\dd t^{2}+h^{2/3}\left(\dd x_{1}^{2}+\cdots+\dd x_{4}^{2}\right)\,,\quad e^{\sigma}=h^{2/\sqrt{6}}\,,\\
    F^{1}_{(2)}&=\sqrt{2}\partial_{a}h^{-1}\dd t\wedge\dd x^{a}\,,\quad\partial^{a}\partial_{a}h\left(x_{1},\cdots,x_{4}\right)=0\,,\ a,b=1,2,\cdots4\,,
  \end{aligned}
  \label{eq:5d-bh}
\end{equation}
coupling to the $F$-flux, and another class of solutions
\begin{equation}
  \begin{aligned}
    \dd s_{5}^{2}&=-h^{-2/3}\dd t^{2}+h^{1/3}\left(\dd x_{1}^{2}+\cdots+\dd x_{4}^{2}\right)\,,\quad e^{\sigma}=h^{-2/\sqrt{6}}\,,\\
    G_{(2)}&=h^{-2}\partial_{a}h\,\dd t\wedge\dd x^{a}\,,\quad\partial^{a}\partial_{a}h\left(x_{1},\cdots,x_{4}\right)=0\,,\ a,b=1,2,\cdots4\,,
  \end{aligned}
  \label{eq:5d-bh-g}
\end{equation}
coupling to the $G$-flux. 
Similarly, we can construct black-string solutions, which magnetically couple to the fluxes.

\section{Supersymmetry projections of BPS-branes}\label{app:susy}
The preserved supersymmetry of BPS-brane solutions is encoded in projectors composed of gamma matrices. The projections on the Killing spinors of the BPS-branes are \cite{Gauntlett:1997cv}
\begin{equation}
  \begin{aligned}
    \text{M2-brane:}&\qquad \epsilon = \Gamma_{\underline{012}}\epsilon\,,\\
    \text{M5-brane:}&\qquad \epsilon = \Gamma_{\underline{012345}}\epsilon\,,\\
    \text{F1-string:}&\qquad \epsilon_L = \Gamma_{\underline{01}}\epsilon_L\,,\qquad\quad\ \, \epsilon_R = -\Gamma_{\underline{01}}\epsilon_R\,,\\
    \text{IIA NS5-brane:}&\qquad \epsilon_L = \Gamma_{\underline{012345}}\epsilon_L\,,\qquad \epsilon_R = \Gamma_{\underline{012345}}\epsilon_R\,,\\
    \text{IIB NS5-brane:}&\qquad \epsilon_L = \Gamma_{\underline{012345}}\epsilon_L\,,\qquad \epsilon_R = -\Gamma_{\underline{012345}}\epsilon_R\,,\\
    \text{D}p\text{-brane}\text{:}&\qquad \epsilon_L = \Gamma_{\underline{01\cdots p}}\epsilon_R\,.
  \end{aligned}
  \label{eq:projections-BPS}
\end{equation}
In the 10d theories, we act on left-handed and right-handed spinors as defined in \eqref{eq:IIA-spinor-separate} and \eqref{eq:IIB-spinor-separate}.

When a brane is added to a system, it imposes a projection condition according to \eqref{eq:projections-BPS} on the Killing spinor. In general, such a projection breaks half of the supersymmetry. However, if the new projector anti-commutes with the existing ones, it breaks all the supersymmetry. On the other hand, if the projection of the new brane is implied by the existing ones, the brane can be added without breaking any additional supersymmetry. 

\section{Fermionic embedding ansatz and proof}\label{app:fermion-consistency}

\subsection{Fermionic embedding ansatz}\label{fermion-embedding}
The key to constructing the fermionic ansatz for embedding supergravity on a skeleton brane is to promote the Killing spinor of the skeleton brane to a local supersymmetry parameter.
As a first step, we decompose the Killing spinor into a tensor product of spinors on the worldvolume and transverse spaces
\begin{equation}
  \epsilon = \varepsilon^A\otimes\eta_A\,.
\end{equation}
Here, $\varepsilon^A$ are constant Grassmann-odd spinors labelled by $A$, and they satisfy the BPS projection conditions of the skeleton brane. All spacetime dependence is encoded in the transverse-space Grassmann-even spinors $\eta_A$\footnote{The Grassmann-even spinor $\eta_A$ is chosen to make sure the higher-dimensional spinor is Grassmann-odd. In this paper, the Grassmann parity does not play a huge r\^{o}le as we only consider the equations of motion and supersymmetry transformations to linear order in fermions. However, it would become important when considering the truncations to the full-fermionic level \cite{Lin:2024eqq} or quantisation of the theory.}. The index $A$ on $\eta_A$ can be interpreted in two equivalent ways: it either labels the linearly independent spinors forming a basis of the transverse Killing spinor space, or it serves as a spinor index of the tangent space of a point in the transverse space. The fermionic ansatz is then constructed by promoting the constant worldvolume spinors $\varepsilon^A$ to local functions acting as supersymmetry generators in lower dimensions, while keeping the transverse Killing spinors $\eta_A$ unchanged. Generically, the ten-dimensional gravitino decomposes into a linear combination of the lower-dimensional gravitino and dilatino (except for the  D3-brane case).

\

To get the fermionic ansatz for 5d $\mathcal{N}=4$ supergravity, we need to generalise the Killing spinor of the D4-brane \eqref{eq:D4-spinor-factorisation}. The idea is to promote the constant Killing spinors tangent to the worldvolume to 5d supersymmetry transformation parameters. This procedure is complicated due to the warping factor, \ie $\exp(\frac{5}{8\sqrt{6}}\sigma)$, but a sufficiently general ansatz allows us to determine
\begin{equation}
  \begin{aligned}
    \hat{\Psi}_{\mu}=&e^{\frac{5}{32\sqrt{6}}\sigma}\left\{ \left(\begin{array}{c}
    1\\
    -i
    \end{array}\right)\otimes\left(\psi_{\mu}^{i}+\frac{5i}{16\sqrt{3}}\gamma_{\mu}\chi^{i}\right)\otimes\eta_{i}\right\} \,,\\
    \hat{\Psi}_{I}=&-\frac{\sqrt{3}}{16}H_{0}^{\frac{3}{16}}e^{-\frac{5}{16\sqrt{6}}\sigma}e_I{}^{\underline{J}}(\sigma_3\otimes\mathbf{1}\otimes\Sigma_{\underline{J}})\left\{\left(\begin{array}{c}
    1\\
    -i
    \end{array}\right)\otimes\chi^{i}\otimes\eta_{i}\right\}\,,\\
    \hat{\lambda}=&-\frac{\sqrt{3}}{2}H_{0}^{\frac{3}{16}}e^{-\frac{5}{32\sqrt{6}}\sigma}\left(\begin{array}{c}
    1\\
    -i
    \end{array}\right)\otimes\chi^{i}\otimes\eta_{i}\,,\qquad\hat{\epsilon}=e^{\frac{5}{32\sqrt{6}}\sigma}\left(\begin{array}{c}
    1\\
    -i
    \end{array}\right)\otimes\varepsilon^{i}\left(x\right)\otimes\eta_{i}\,.
  \end{aligned}
  \label{eq:5dN4-embedding-fermion}
\end{equation}
The gamma matrices are given in \eqref{eq:IIA-gamma-factorise}. The f\"unfbein $e_I{}^{\underline{J}}=H_0^{\frac{5}{16}}e^{-\frac{3}{16\sqrt{6}}\sigma}\delta_I^{\underline{J}}$ and its inverse $e_{\underline{J}}{}^{I}=(e_I{}^{\underline{J}})^{-1}$ relate Einstein and Lorentz indices in the transverse space \eqref{eq:5dN4-bosonic-ansatz}. Both the worldvolume and the transverse-space spinors are symplectic Majorana spinors transforming under the $\mathbf{4}$ representation of $USp(4)$. $H_0$ is the harmonic function of the skeleton brane, \ie a D4-brane. 

The proof of consistency is similar to the 6d $\mathcal{N}=(1,1)$ case given in the next subsection. The first step is to substitute \eqref{eq:5dN4-embedding-fermion} into the supersymmetry transformation \eqref{eq:IIA-susy-bosonic} in the bosonic background \eqref{eq:5dN4-bosonic-ansatz}. The relation 
\begin{equation}
  \Sigma_{\underline{I}}\eta_{i}=2\,\eta_{j}\left(L_{\underline{I}}\right)_{\ i}^{j}\,
  \label{eq:5dN4-susy-key}
\end{equation}
satisfied by the local intertwiner between the $\mathbf{4}$ representations of $USp(4)$ and $SO(5)$, Eq.\eqref{eq:USp4-SO5-intertwiner}, is vital to split the supersymmetry transformations into worldvolume and transverse spaces. After a straightforward calculation, the IIA supersymmetry transformations \eqref{eq:IIA-susy-bosonic} reduce to the 5d ones \eqref{eq:5DN4-bosonic-susy}. 

\

The fermionic ansatz for the 6d $\mathcal{N}=(1,1)$ supergravity is again generated by the NS5-brane Killing spinor \eqref{eq:KS-NS5}. A similar approach to the case above gives
\begin{equation}
  \begin{aligned}
    \hat{\Psi}_{\mu}=&e^{-\frac{\sqrt{2}}{16}\varphi}\left\{ \left(\psi_{\mu}^{A}\otimes\eta_{A}+i\psi_{\mu}^{\dot{A}}\otimes\eta_{\dot{A}}\right)+\frac{\sqrt{2}}{4}\gamma_{\mu}\left(\chi^{A}\otimes\eta_{A}+i\chi^{\dot{A}}\otimes\eta_{\dot{A}}\right)\right\} \,,\\
    \hat{\Psi}_{I}=&-\frac{1}{2\sqrt{2}}H_{0}^{\frac{1}{8}}e^{\frac{1}{8\sqrt{2}}\varphi}e_I{}^{\underline{J}}\left(\gamma^{(6)}\otimes\Sigma_{\underline{J}}\right)\left(\chi^{A}\otimes\eta_{A}+i\chi^{\dot{A}}\otimes\eta_{\dot{A}}\right)\,,\\\hat{\lambda}=&\sqrt{2}H_{0}^{\frac{1}{8}}e^{\frac{\sqrt{2}}{16}\varphi}\left(\chi^{A}\otimes\eta_{A}+i\chi^{\dot{A}}\otimes\eta_{\dot{A}}\right)\,,\\\hat{\epsilon}=&e^{-\frac{\sqrt{2}}{16}\varphi}\varepsilon\,,\qquad\varepsilon=\varepsilon^{A}\otimes\eta_{A}+i\varepsilon^{\dot{A}}\otimes\eta_{\dot{A}}\,.
  \end{aligned}
  \label{eq:6d11-fermionic-ansatz}
\end{equation}
Here, the gamma matrices are chosen as in \eqref{eq:IIB-gamma-factorise}. The vierbein $e_I{}^{\underline{J}}=H_0^{\frac{3}{8}}e^{\frac{1}{4\sqrt{2}}\sigma}\delta_I^{\underline{J}}$ and its inverse $e_{\underline{J}}{}^{I}=(e_I{}^{\underline{J}})^{-1}$ relate Einstein and Lorentz indices in the transverse space \eqref{eq:6d11-bosonic-ansatz}. The 6d spinors are given in \eqref{eq:6d-spinors-separation} while the 4d spinors are given in \eqref{eq:5d-transverse-spinor}. $H_0$ is the harmonic function of the skeleton brane, \ie an NS5-brane. The proof of consistency is given in the next subsection.

\subsection{Consistency of the 6d $\mathcal{N}=(1,1)$ embedding ansatz}\label{consistency}
We now prove that the bosonic ansatz \eqref{eq:6d11-bosonic-ansatz} is compatible with the bosonic equations of motion and Bianchi identities. We first take note of some useful identities 
\begin{equation}
  \begin{aligned}
    *\tilde{F}_{(5)}=&\frac{1}{\sqrt{2}}\left[e^{-\frac{\sqrt{2}}{2}\varphi}*_{g}F_{(2)}^{I}\wedge\dd y^{I}-\frac{1}{3!}H_{0}\varepsilon_{I_{1}\cdots I_{4}}F_{(2)}^{I_{1}}\wedge\dd y^{I_{2}}\wedge\dd y^{I_{3}}\dd y^{I_{4}}\right]\,,\\
    *\hat{H}_{(3)}=&H_{0}^{-\frac{3}{2}}e^{-\sqrt{2}\varphi}\partial_{I}H_{0}\dd y^{I}\wedge\text{vol}_{g}- H_{0}^{\frac{3}{2}}e^{\frac{\sqrt{2}}{2}\varphi}*_{g}H_{(3)}\wedge\text{vol}_{\delta}\,,\\
    *\tilde{F}_{(3)}=&-\frac{1}{\sqrt{2}}\left(\frac{1}{3!}H_{0}^{\frac{1}{2}}\varepsilon_{I_{1}I_{2}I_{3}I_{4}}*_{g}F_{(2)}^{I_{1}}\left(x\right)\wedge\dd y^{I_{2}}\wedge\cdots\wedge\dd y^{I_{4}}\right)\,.
  \end{aligned}
  \label{eq:Proofing-identity}
\end{equation}
The Bianchi identity for $\hat{H}_{(3)}$ reduces to the Bianchi identity for $H_{(3)}$ in 6d
\begin{equation}
  \dd H_{(3)}=0\,,
\end{equation}
because $H_{0}$ is a harmonic function. The Bianchi identity for $\tilde{F}_{(3)}$ directly reduces to the Bianchi identity for $F_{(2)}^{I} $
\begin{equation}
  \dd F_{(2)}^I=0\,.
\end{equation}
In the Bianchi identity for $\tilde{F}_{(5)}$, the LHS reduces to 
\begin{equation}
    \dd\tilde{F}_{(5)}=-\frac{1}{\sqrt{2}}\partial_{I}H_{0}F_{(2)}^{I}\wedge\text{vol}_{\delta}+\frac{1}{\sqrt{2}}\dd\left(e^{-\frac{\sqrt{2}}{2}\varphi}*_{g}F_{(2)}^{I}\right)\wedge\dd y^{I}\,,
\end{equation}
where we used the Bianchi identity for $F_{(2)}^{I}$, while the RHS reads
\begin{equation}
  -\tilde{F}_{(3)}\wedge \hat{H}_{(3)}=-\frac{1}{\sqrt{2}}\partial_{I}H_{0}F_{(2)}^{I}\left(x\right)\wedge\text{vol}_{\delta}+\frac{1}{\sqrt{2}}H_{(3)}\left(x\right)\wedge F_{(2)}^{I}\left(x\right)\wedge\dd y^{I}\,.
\end{equation}
We get the equation of motion of $F_{(2)}^{I}$ in 6d 
\begin{equation}
  \dd\left(e^{-\frac{\sqrt{2}}{2}\varphi}*_{g}F_{(2)}^{I}\right)- H_{(3)}\left(x\right)\wedge F_{(2)}^{I}\left(x\right)=0\,,
\end{equation}
after combining them.
For the individual terms in the dilaton equation, we may use the identities \eqref{eq:Proofing-identity} to infer 
\begin{equation}
  \begin{aligned}
    \dd*\dd\hat{\Phi} =&-\frac{1}{2}e^{-\frac{\sqrt{2}}{2}\varphi}H_{0}^{-2}\partial_{I}H_{0}\partial_{I}H_{0}\text{vol}_{\delta}\wedge\text{vol}_{g}-\frac{1}{\sqrt{2}}H_{0}\dd*\dd\varphi\wedge\text{vol}_{\delta}\,,\\
    \frac{1}{2}e^{\hat{\Phi}}\tilde{F}_{(3)}\wedge*\tilde{F}_{(3)}=&\frac{1}{4}H_{0}e^{-\frac{\varphi}{\sqrt{2}}}\delta_{IJ}F_{(2)}^{I}\left(x\right)\wedge*_{g}F^{J}\left(x\right)\wedge\text{vol}_{\delta}\,,\\
    -\frac{1}{2}e^{-\hat{\Phi}}\hat{H}_{(3)}\wedge*\hat{H}_{(3)}=&-\frac{1}{2}H_{0}^{-2}e^{-\frac{\varphi}{\sqrt{2}}}\left(\partial_{I}H_{0}\partial_{I}H_{0}\text{vol}_{\delta}\wedge\text{vol}_{g}\right)\\
    &-\frac{1}{2}H_{0}e^{\sqrt{2}\varphi}H_{(3)}\left(x\right)\wedge*_{g}H_{(3)}\wedge\text{vol}_{\delta}\,.\\
  \end{aligned}
\end{equation}
After combining these, we get the scalar equation in 6d 
\begin{equation}
  0=\dd*\dd\varphi+\frac{1}{2\sqrt{2}}e^{-\frac{\varphi}{\sqrt{2}}}\delta_{IJ}F_{(2)}^{I}\left(x\right)\wedge*_{g}F_{(2)}^{J}\left(x\right)-\frac{\sqrt{2}}{2}e^{\sqrt{2}\varphi}H_{(3)}\left(x\right)\wedge*_{g}H_{(3)}\,.
\end{equation}
The axion equation vanishes trivially. Given the self-duality of $\tilde{F}_{(5)}$, which is easy to see from \eqref{eq:Proofing-identity}, the equation of motion becomes a Bianchi identity and results in the $F_{(2)}^{I}$ equation of motion in 6d. In the  equation of motion for $\tilde{F}_{(3)}$, the LHS is 
\begin{equation}
  \begin{aligned}
    \dd\left(e^{\hat{\Phi}}*\tilde{F}_{(3)}\right)=&-\frac{1}{\sqrt{2}}e^{-\frac{\varphi}{\sqrt{2}}}\partial_{I}H_{0}*_{g}F_{(2)}^{I}\left(x\right)\wedge\text{vol}_{y}\\
    &-\frac{1}{\sqrt{2}}\frac{1}{3!}H_{0}\varepsilon_{I_{1}I_{2}I_{3}I_{4}}\dd\left(e^{-\frac{\varphi}{\sqrt{2}}}*_{g}F_{(2)}^{I_{1}}\left(x\right)\right)\wedge\dd y^{I_{2}}\wedge\cdots\wedge\dd y^{I_{4}}\,,
  \end{aligned}
\end{equation}
while the RHS is
\begin{equation}
  \begin{aligned}
    \tilde{F}_{(5)}\wedge \hat{H}_{(3)}=&-\frac{1}{\sqrt{2}}e^{-\frac{\sqrt{2}}{2}\varphi}\partial_{I}H_{0}*_{g}F_{(2)}^{I}\wedge\text{vol}_{y}\\
    &+\frac{1}{\sqrt{2}}\left[\frac{1}{3!}H_{0}\varepsilon_{I_{1}\cdots I_{4}}F_{(2)}^{I_{1}}\wedge\dd y^{I_{2}}\wedge\dd y^{I_{3}}\wedge\dd y^{I_{4}}\right]\wedge H_{(3)}\left(x\right)\,.
  \end{aligned}
\end{equation}
These combine to give the $F_{(2)}^{I}$ equation of motion in 6d. In the equation of motion of $\hat{H}_{(3)}$, the LHS reads
\begin{equation}
  \dd\left(e^{-\hat{\Phi}}*\hat{H}_{(3)}\right) =-H_{0}\dd\left(e^{\sqrt{2}\varphi}*_{g}H_{(3)}\right)\wedge\text{vol}_{y}\,
\end{equation}
and the RHS is 
\begin{equation}
  -\tilde{F}_{(5)}\wedge \tilde{F}_{(3)}=\frac{1}{2}H_{0}\delta_{IJ}F_{(2)}^{I}\wedge F_{(2)}^{J}\left(x\right)\wedge\text{vol}_{y}\,.
\end{equation}
Hence, the $\hat{H}_{(3)}$ equation of motion reduces to 
\begin{equation}
  \dd\left(e^{\sqrt{2}\varphi}*_{g}H_{(3)}\right)+\frac{1}{2}\delta_{IJ}F_{(2)}^{I}\wedge F_{(2)}^{J}\left(x\right)=0\,,
\end{equation}
which is the 6d $H_{(3)}$ equation of motion. 

Finally we have to consider Einstein's equation. We determine the zehnbein to be
\begin{equation}
  \hat{e}^{\underline{\mu}}=H_{0}^{-\frac{1}{8}}e^{-\frac{\sqrt{2}}{8}\varphi}e^{\underline{\mu}}\,,\qquad \hat{e}^{\underline{I}}=H_{0}^{\frac{3}{8}}e^{\frac{\sqrt{2}}{8}\varphi}\dd y^{I}\,.
\end{equation}
After a tedious calculation, one may use the Cartan structure equations and determine the Ricci tensor  
\begin{equation}
  \begin{aligned}
    \hat{R}_{\mu\nu}=&R_{\mu\nu}-\frac{1}{4}\partial_{\mu}\varphi\partial_{\nu}\varphi+\frac{\sqrt{2}}{8}\nabla_{\rho}\left(g^{\rho\eta}\partial_{\eta}\varphi\right)g_{\mu\nu}+\frac{1}{8}g_{\mu\nu}H_{0}^{-1}e^{-\frac{\sqrt{2}}{2}\varphi}\partial_{K}H_{0}^{-1}\partial_{K}H_{0}\,,\\
    \hat{R}_{IJ}=&-\frac{\sqrt{2}}{8}H_{0}e^{\frac{\sqrt{2}}{2}\varphi}\nabla_{\rho}\left(\partial_{\sigma}\varphi g^{\rho\sigma}\right)\delta_{IJ}+\frac{3}{8}\partial_{J}H_{0}^{-1}\partial_{I}H_{0}-\frac{3}{8}\delta_{IJ}\partial_{K}H_{0}^{-1}\partial_{K}H_{0}\,,\\
    \hat{R}_{\mu I}=&-\frac{\sqrt{2}}{8}H_{0}^{-1}\partial_{I}H_{0}\partial_{\mu}\varphi\,.\\
  \end{aligned}
\end{equation}
After another straightforward calculation, one can derive the RHS of the Einstein equation \eqref{eq:IIB-bosonic-eom} 
\begin{equation}
  \begin{aligned}
    \text{RHS}_{\mu\nu}=&\frac{1}{4}\partial_{\mu}\varphi\partial_{\nu}\varphi+\frac{1}{4}e^{\sqrt{2}\varphi}\left(H_{\mu\rho\sigma}H_{\nu}^{\ \rho\sigma}-\frac{1}{12}g_{\mu\nu}H_{\rho\sigma\lambda}H^{\rho\sigma\lambda}\right)\\
    &+\frac{1}{2}e^{-\frac{\varphi}{\sqrt{2}}}\delta_{IJ}\left(F_{\mu\rho I}^{I}F_{\nu}^{J\rho}-\frac{3}{16}g_{\mu\nu}F_{\rho\sigma}^{I}F^{J\rho\sigma}\right)+\frac{1}{8}H^{-1}e^{-\frac{\varphi}{\sqrt{2}}}g_{\mu\nu}\partial_{K}H_{0}^{-1}\partial_{K}H_{0}\,,\\
    \text{RHS}_{IJ}=&\left(\frac{1}{32}H_{0}F_{\rho\sigma}^{K}F^{K\rho\sigma}-\frac{1}{48}H_{0}e^{\frac{3\sqrt{2}}{2}\varphi}H_{\rho\sigma\lambda}H^{\rho\sigma\lambda}\right)\delta_{IJ}+\frac{3}{8}\partial_{I}H_{0}^{-1}\partial_{J}H_{0}-\frac{3}{8}\delta_{IJ}\partial_{M}H_{0}^{-1}\partial_{M}H_{0}\,,\\
    \text{RHS}_{\mu I}=&-\frac{1}{4\sqrt{2}}H_{0}^{-1}\partial_{I}H_{0}\partial_{\mu}\varphi\,.
  \end{aligned}
\end{equation}
We can see that the $\mu I$ and $IJ$ components of the Einstein equation are satisfied trivially. The $\mu\nu$ components of the Einstein equation reduce to the 6d Einstein equation 
\begin{equation}
  \begin{aligned}
    R_{\mu\nu}=&\frac{1}{2}\partial_{\mu}\varphi\partial_{\nu}\varphi+\frac{1}{2}e^{-\frac{\varphi}{\sqrt{2}}}\delta_{IJ}\left(F_{\mu\rho}^{I}F_{\nu}^{J\rho}-\frac{1}{8}g_{\mu\nu}F_{\rho\sigma}^{I}F^{J\rho\sigma}\right)\\&+e^{\sqrt{2}\varphi}\frac{1}{4}\left(H_{\mu\rho\sigma}H_{\nu}^{\ \rho\sigma}-\frac{1}{6}g_{\mu\nu}H_{\rho\sigma\lambda}H^{\rho\sigma\lambda}\right)\,.
  \end{aligned}
\end{equation}

This concludes the proof of consistency at the bosonic level.

\

Let us now proceed with proving consistency of the fermionic ansatz \eqref{eq:6d11-fermionic-ansatz}. Substitute the embedding ansatz \eqref{eq:6d11-bosonic-ansatz} and \eqref{eq:6d11-fermionic-ansatz} into the type IIB supersymmetry transformation \eqref{eq:IIB-susy}. After a straightforward calculation, the dilatino transformation gives 
\begin{equation}
  \begin{aligned}
    \delta\hat{\lambda}=&\sqrt{2}H_{0}^{\frac{1}{8}}e^{\frac{\sqrt{2}}{16}\varphi}\left(\delta\chi^{A}\otimes\eta_{A}+i\delta\chi^{\dot{A}}\otimes\eta_{\dot{A}}\right)\\
    =&\sqrt{2}H_{0}^{\frac{1}{8}}e^{\frac{\sqrt{2}}{16}\varphi}\left\{ -\frac{1}{4}\partial_{\mu}\varphi\left(\gamma^{\mu}\otimes\mathbf{1}\right)\left(\varepsilon^{A}\otimes\eta_{A}+i\varepsilon^{\dot{A}}\otimes\eta_{\dot{A}}\right)\right.\\&+\left.\left(\frac{1}{24\sqrt{2}}e^{\frac{\varphi}{\sqrt{2}}}H_{\mu\nu\rho}\left(\gamma^{\mu\nu\rho}\otimes\mathbf{1}\right)+\frac{i}{16}e^{-\frac{\varphi}{2\sqrt{2}}}F_{\mu\nu}^{I}\delta_I^{\underline{J}}\left(\gamma^{\mu\nu}\gamma^{(6)}\otimes\Sigma_{\underline{J}}\right)\right)\left(\varepsilon^{A}\otimes\eta_{A}-i\varepsilon^{\dot{A}}\otimes\eta_{\dot{A}}\right)\right\}\,.
  \end{aligned}
  \label{eq:proof-dilaton}
\end{equation}
Here we explicitly evaluated the vierbein $e_{I}{}^{\underline{J}}$ and pulled out an overall factor.
Similarly, the $\mu$-component of the gravitino transformation gives 
\begin{equation}
  \begin{aligned}
    \delta\hat{\Psi}_{\mu}=&e^{-\frac{\sqrt{2}}{16}\varphi}\delta\left\{ \left(\psi_{\mu}^{A}\otimes\eta_{A}+i\psi_{\mu}^{\dot{A}}\otimes\eta_{\dot{A}}\right)+\frac{\sqrt{2}}{4}\gamma_{\mu}\left(\chi^{A}\otimes\eta_{A}+i\chi^{\dot{A}}\otimes\eta_{\dot{A}}\right)\right\} \\
    =&e^{-\frac{\sqrt{2}}{16}\varphi}\left\{ D_{\mu}\hat{\epsilon}+\frac{i}{16\sqrt{2}}e^{-\frac{\varphi}{2\sqrt{2}}}F_{\rho\sigma}^{I}\delta_I^{\underline{J}}\left(\left(\gamma_{\mu}^{\ \rho\sigma}-6\delta_{\mu}^{\rho}\gamma^{\sigma}\right)\gamma^{(6)}\otimes\Sigma^{\underline{I}}\right)\hat{\epsilon}^{c}\right.\\
    &\left.-\frac{1}{48}e^{\frac{\varphi}{\sqrt{2}}}H_{\nu\rho\sigma}\left[\left(\gamma_{\mu}^{\ \nu\rho\sigma}-3\delta_{\mu}^{\nu}\gamma^{\rho\sigma}\right)\otimes\mathbf{1}\right]\hat{\epsilon}^{c}\right\} \\
    &+e^{-\frac{\sqrt{2}}{16}\varphi}\frac{\sqrt{2}}{4}\left(\gamma_{\mu}\otimes\mathbf{1}\right)\left\{-\frac{1}{4}\partial_{\lambda}\varphi\left(\gamma^{\lambda}\otimes\mathbf{1}\right)\hat{\epsilon}\right.\\
    &+\left.\left(\frac{1}{24\sqrt{2}}e^{\frac{\varphi}{\sqrt{2}}}H_{\rho\sigma\lambda}\left(\gamma^{\rho\sigma\lambda}\otimes\mathbf{1}\right)+\frac{i}{16}e^{-\frac{\varphi}{2\sqrt{2}}}F_{\rho\sigma}^{I}\delta_I^{\underline{J}}\left(\gamma^{\rho\sigma}\gamma^{(6)}\otimes\Sigma_{\underline{J}}\right)\right)\hat{\epsilon}^{c}\right\}\,.
  \end{aligned}\label{eq:gravitino}
\end{equation}
The $I$-component of the gravitino transformation gives 
\begin{equation}
  \begin{aligned}
    \delta\hat{\Psi}_{I}=&-\frac{1}{2\sqrt{2}}H_{0}^{\frac{1}{8}}e^{\frac{1}{8\sqrt{2}}\varphi}e_I{}^{\underline{J}}\left(\gamma^{(6)}\otimes\Sigma_{\underline{J}}\right)\delta\left(\chi^{A}\otimes\eta_{A}+i\chi^{\dot{A}}\otimes\eta_{\dot{A}}\right)\\
    =&-\frac{1}{2\sqrt{2}}H_{0}^{\frac{1}{8}}e^{\frac{1}{8\sqrt{2}}\varphi}e_I{}^{\underline{J}}\left(\gamma^{(6)}\otimes\Sigma_{\underline{J}}\right)
    \left\{-\frac{1}{4}\partial_{\nu}\varphi\left(\gamma^{\nu}\otimes\mathbf{1}\right)\hat{\epsilon}\right.\\
    &+\left.\left(\frac{1}{24\sqrt{2}}\left(e^{\frac{1}{\sqrt{2}}\varphi}H_{\mu\nu\rho}\left(\gamma^{\mu\nu\rho}\otimes\mathbf{1}\right)\right)+\frac{i}{16}e^{-\frac{1}{2\sqrt{2}}\varphi}F_{\mu\nu}^{K}\delta_K^{\underline{L}}\left(\gamma^{\mu\nu}\gamma^{(6)}\otimes\Sigma_{\underline{L}}\right)\right)\hat{\epsilon}^{c}\right\}\ .
  \end{aligned}
  \label{eq:proof-gravitino}
\end{equation}
We can see that the summation over $F^I_{(2)}$ and $\Sigma^{\underline{J}}$ prevents us from decoupling the transformations (\ref{eq:proof-dilaton}-\ref{eq:proof-gravitino}) into 6d and 4d parts. 

Pauli matrices naturally serve as the 
intertwiner between $Spin(4)$ and $SU(2)\times SU(2)$, 
\begin{equation}
  \Sigma^{\underline{0}}=\left(\begin{array}{cc}
   & \delta_{\ \dot{B}}^{A}\\
  \delta_{\ B}^{\dot{A}}
  \end{array}\right)\,,\qquad\Sigma^{\underline{a}}=\left(\begin{array}{cc}
   & \left(\sigma^{a}\right)_{\ \dot{B}}^{A}\\
  \left(\sigma^{a}\right)_{\ B}^{\dot{A}}
  \end{array}\right)\,.
  \label{eq:6d-intertwiner}
\end{equation}
Here, we use $(0,a)$ indices instead of the indices $I,J$ as in \eqref{eq:6d-susy}, where $a$ is a fundamental $SO(3)\simeq SU(2)\subset SO(4)$ index.
From this one obtains the key relations
\begin{equation}
  \begin{aligned}
    &\varepsilon^{A}\otimes\Sigma^{\underline{0}}\eta_{A}	=\varepsilon^{\dot{A}}\otimes\eta_{\dot{A}}\,,\qquad\qquad\ \ \, 
    \varepsilon^{\dot{A}}\otimes\Sigma^{\underline{a}}\eta_{\dot{A}}	=\varepsilon^{A}\otimes\eta_{A}\,,\\
    &\varepsilon^{A}\otimes\Sigma^{\underline{a}}\eta_{A}	=\left(\sigma^{a}\right)_{\ C}^{\dot{A}}\varepsilon^{C}\otimes\eta_{\dot{A}}\,,\qquad 
  \varepsilon^{\dot{A}}\otimes\Sigma^{\underline{a}}\eta_{\dot{A}}	=\left(\sigma^{a}\right)_{\ \dot{C}}^{A}\varepsilon^{\dot{C}}\otimes\eta_{A}\,.
  \end{aligned}
  \label{eq:6d-key-relation}
\end{equation}
Then, we need to make a projection in the transverse directions and use 
the relation \eqref{eq:6d-key-relation} to decouple the 6d and 4d parts of the 10d spinor. One obtains 
\begin{equation}
  \begin{aligned}
    \delta\chi^{A}=&-\frac{1}{4}\partial_{\mu}\varphi\gamma^{\mu}\varepsilon^{A}+\frac{1}{24\sqrt{2}}e^{\frac{\varphi}{\sqrt{2}}}H_{\mu\nu\rho}\gamma^{(6)}\gamma^{\mu\nu\rho}\varepsilon^{A}\\
    &\quad-\frac{1}{16}e^{-\frac{\varphi}{2\sqrt{2}}}F_{\mu\nu}^{0}\gamma^{(6)}\gamma^{\mu\nu}\varepsilon^{A}+\frac{1}{16}e^{-\frac{\varphi}{2\sqrt{2}}}F_{\mu\nu}^{a}\gamma^{\mu\nu}\left(\sigma^{a}\right)_{\ \dot{C}}^{A}\varepsilon^{\dot{C}}\,,\\\delta\chi^{\dot{A}}=&-\frac{1}{4}\partial_{\mu}\varphi\gamma^{\mu}\varepsilon^{\dot{A}}+\frac{1}{24\sqrt{2}}e^{\frac{\varphi}{\sqrt{2}}}H_{\mu\nu\rho}\gamma^{(6)}\gamma^{\mu\nu\rho}\varepsilon^{\dot{A}}\\
    &\quad-\frac{1}{16}e^{-\frac{\varphi}{2\sqrt{2}}}F_{\mu\nu}^{I}\gamma^{(6)}\gamma^{\mu\nu}\varepsilon^{\dot{A}}-\frac{1}{16}e^{-\frac{\varphi}{2\sqrt{2}}}F_{\mu\nu}^{a}\gamma^{\mu\nu}\left(\sigma^{a}\right)_{\ C}^{\dot{A}}\varepsilon^{C}\,,
  \end{aligned}
\end{equation}
from both the transformations of the dilatino and of the $I$-component of the gravitino. After combining them with \eqref{eq:6d-spinors-separation}  we have 
\begin{equation}
  \begin{aligned}
    \delta\chi^i=&-\frac{1}{4}\partial_{\mu}\varphi\gamma^{\mu}\varepsilon^i+\frac{1}{24\sqrt{2}}e^{\frac{\varphi}{\sqrt{2}}}H_{\mu\nu\rho}\gamma^{(6)}\gamma^{\mu\nu\rho}\varepsilon^i\\
    &\quad-\frac{1}{16}e^{-\frac{\varphi}{2\sqrt{2}}}\left[F_{\mu\nu}^{0}\gamma^{(6)}\delta^i_j+iF_{\mu\nu}^{a}(\sigma^a)^i{}_j\right]\gamma^{\mu\nu}\varepsilon^j\ ,
  \end{aligned}
\end{equation}
which is related to the 6d dilatino transformation \eqref{eq:6d-susy} by a $SU(2)$ transformation. Similarly, from \eqref{eq:gravitino} we get the gravitino transformation 
\begin{equation}
  \begin{aligned}
    \delta\psi_{\mu}^{i}=&D_{\mu}\varepsilon^i+\frac{1}{48}e^{\frac{\varphi}{\sqrt{2}}}H_{\nu\rho\sigma}\gamma^{(6)}\left(\gamma_{\mu}^{\ \nu\rho\sigma}-3\delta_{\mu}^{\nu}\gamma^{\rho\sigma}\right)\varepsilon^i\\
    &+\frac{1}{16\sqrt{2}}e^{-\frac{\varphi}{2\sqrt{2}}}\left[F_{\rho\sigma}^{0}\gamma^{(6)}\delta_j^i-iF_{\rho\sigma}^{a}(\sigma^a)^i{}_j\right]\left(\gamma_{\mu}^{\ \rho\sigma}-6\delta_{\mu}^{\rho}\gamma^{\sigma}\right)\varepsilon^j\,,
  \end{aligned}
\end{equation}
 related to the 6d gravitino transformation \eqref{eq:6d-susy} by an $SU(2)$ transformation.

This concludes the proof of consistency at the fermionic level.

\end{appendices}

\bibliographystyle{JHEP}
\bibliography{Intersecting_branes}

@article{Duff:1990xz,
    author = "Duff, M. J. and Stelle, K. S.",
    title = "{Multi-membrane solutions of D = 11 supergravity}",
    reportNumber = "CTP-TAMU-72-90, IMPERIAL-TP-89-90-34",
    doi = "10.1201/9781482268737-12",
    journal = "Phys. Lett. B",
    volume = "253",
    pages = "113--118",
    year = "1991"
}

@article{Gueven:1992hh,
    author = "Gueven, Rahmi",
    title = "{Black p-brane solutions of D = 11 supergravity theory}",
    doi = "10.1201/9781482268737-16",
    journal = "Phys. Lett. B",
    volume = "276",
    pages = "49--55",
    year = "1992"
}

@book{Blumenhagen:2013fgp,
    author = {Blumenhagen, Ralph and L\"ust, Dieter and Theisen, Stefan},
    title = "{Basic concepts of string theory}",
    doi = "10.1007/978-3-642-29497-6",
    isbn = "978-3-642-29496-9",
    publisher = "Springer",
    address = "Heidelberg, Germany",
    series = "Theoretical and Mathematical Physics",
    year = "2013"
}

@article{Tseytlin:1996bh,
    author = "Tseytlin, Arkady A.",
    title = "{Harmonic superpositions of M-branes}",
    eprint = "hep-th/9604035",
    archivePrefix = "arXiv",
    reportNumber = "IMPERIAL-TP-95-96-38",
    doi = "10.1201/9781482268737-28",
    journal = "Nucl. Phys. B",
    volume = "475",
    pages = "149--163",
    year = "1996"
}

@article{Gauntlett:1996pb,
    author = "Gauntlett, Jerome P. and Kastor, David A. and Traschen, Jennie H.",
    title = "{Overlapping branes in M theory}",
    eprint = "hep-th/9604179",
    archivePrefix = "arXiv",
    reportNumber = "CALT-68-2055, QMW-PH-96-8, UMHEP-429",
    doi = "10.1016/0550-3213(96)00423-3",
    journal = "Nucl. Phys. B",
    volume = "478",
    pages = "544--560",
    year = "1996"
}

@inproceedings{Gauntlett:1997cv,
    author = "Gauntlett, Jerome P.",
    title = "{Intersecting branes}",
    booktitle = "{APCTP Winter School on Dualities of Gauge and String Theories}",
    eprint = "hep-th/9705011",
    archivePrefix = "arXiv",
    reportNumber = "QMW-PH-97-13, NI-97023",
    doi = "10.1142/9789814447287_0004",
    pages = "146--193",
    month = "5",
    year = "1997"
}

@article{Lu:1996mg,
    author = "Lu, Hong and Pope, C. N. and Stelle, K. S.",
    title = "{Vertical versus diagonal dimensional reduction for p-branes}",
    eprint = "hep-th/9605082",
    archivePrefix = "arXiv",
    reportNumber = "CTP-TAMU-18-96, IMPERIAL-TP-95-96-36",
    doi = "10.1016/S0550-3213(96)90137-6",
    journal = "Nucl. Phys. B",
    volume = "481",
    pages = "313--331",
    year = "1996"
}

@article{Youm:1999ti,
    author = "Youm, Donam",
    title = "{Partially localized intersecting BPS branes}",
    eprint = "hep-th/9902208",
    archivePrefix = "arXiv",
    reportNumber = "CERN-TH-99-48",
    doi = "10.1016/S0550-3213(99)00384-3",
    journal = "Nucl. Phys. B",
    volume = "556",
    pages = "222--246",
    year = "1999"
}

@article{Smith:2002wn,
    author = "Smith, Douglas J.",
    title = "{Intersecting brane solutions in string and M theory}",
    eprint = "hep-th/0210157",
    archivePrefix = "arXiv",
    reportNumber = "DTP-02-67",
    doi = "10.1088/0264-9381/20/9/203",
    journal = "Class. Quant. Grav.",
    volume = "20",
    pages = "R233",
    year = "2003"
}

@article{Lu:2000xc,
    author = "Lu, Hong and Pope, C. N.",
    title = "{Branes on the brane}",
    eprint = "hep-th/0008050",
    archivePrefix = "arXiv",
    reportNumber = "CTP-TAMU-26-00, UPR-899-T",
    doi = "10.1016/S0550-3213(01)00021-9",
    journal = "Nucl. Phys. B",
    volume = "598",
    pages = "492--508",
    year = "2001"
}

@article{Leung:2022nhy,
    author = "Leung, Rahim and Stelle, K. S.",
    title = "{Supergravities on branes}",
    eprint = "2205.13551",
    archivePrefix = "arXiv",
    primaryClass = "hep-th",
    reportNumber = "Imperial/TP/2022/KS/02",
    doi = "10.1007/JHEP09(2022)099",
    journal = "JHEP",
    volume = "09",
    pages = "099",
    year = "2022"
}

@book{becker_becker_schwarz_2006, 
place={Cambridge}, 
title={String Theory and M-Theory: A Modern Introduction},
DOI={10.1017/CBO9780511816086}, 
publisher={Cambridge University Press},
author={Becker, Katrin and Becker, Melanie and Schwarz, John H.}, 
year={2006}}

@article{Dall_Agata_2001,
   title={General matter coupled gauged supergravity in five dimensions},
   volume={612},
   ISSN={0550-3213},
   url={http://dx.doi.org/10.1016/S0550-3213(01)00367-4},
   DOI={10.1016/s0550-3213(01)00367-4},
   number={1-2},
   journal={Nuclear Physics B},
   publisher={Elsevier BV},
   author={Dall'Agata, G. and Herrmann, C. and Zagermann, M.},
   year={2001},
   month=sep, pages={123-150} }

@article{Giani:1984dw,
    author = "Giani, F. and Pernici, M. and van Nieuwenhuizen, P.",
    title = "{GAUGED N=4 d = 6 SUPERGRAVITY}",
    reportNumber = "ITP-SB-84-35",
    doi = "10.1103/PhysRevD.30.1680",
    journal = "Phys. Rev. D",
    volume = "30",
    pages = "1680",
    year = "1984"
}

@article{Romans:1985tw,
    author = "Romans, L. J.",
    editor = "Salam, A. and Sezgin, E.",
    title = "{The F(4) Gauged Supergravity in Six-dimensions}",
    reportNumber = "NSF-ITP-85-137",
    doi = "10.1016/0550-3213(86)90517-1",
    journal = "Nucl. Phys. B",
    volume = "269",
    pages = "691",
    year = "1986"
}

@article{Andrianopoli:2001rs,
    author = "Andrianopoli, Laura and D'Auria, Riccardo and Vaula, Silvia",
    title = "{Matter coupled F(4) gauged supergravity Lagrangian}",
    eprint = "hep-th/0104155",
    archivePrefix = "arXiv",
    doi = "10.1088/1126-6708/2001/05/065",
    journal = "JHEP",
    volume = "05",
    pages = "065",
    year = "2001"
}

@article{Sezgin:2023hkc,
    author = "Sezgin, Ergin",
    title = "{Survey of supergravities}",
    eprint = "2312.06754",
    archivePrefix = "arXiv",
    primaryClass = "hep-th",
    reportNumber = "MI-TH-192",
    month = "12",
    year = "2023"
}

@article{Stelle:1999ljt,
    author = "Stelle, K. S.",
    editor = "Shapiro, M. M. and Silberberg, R. and Wefel, J. P.",
    title = "{BPS Branes in Supergravity}",
    doi = "10.1007/978-94-011-4542-8_12",
    journal = "NATO Sci. Ser. C",
    volume = "530",
    pages = "257--351",
    year = "1999"
}

@article{Campbell:1984zc,
    author = "Campbell, I. C. G. and West, Peter C.",
    title = "{N=2 D=10 Nonchiral Supergravity and Its Spontaneous Compactification}",
    reportNumber = "Print-84-0278 (KING'S COLL)",
    doi = "10.1016/0550-3213(84)90388-2",
    journal = "Nucl. Phys. B",
    volume = "243",
    pages = "112--124",
    year = "1984"
}

@article{Romans:1985ps,
    author = "Romans, L. J.",
    title = "{Gauged $N=4$ Supergravities in Five-dimensions and Their Magnetovac Backgrounds}",
    reportNumber = "NSF-ITP-85-113",
    doi = "10.1016/0550-3213(86)90398-6",
    journal = "Nucl. Phys. B",
    volume = "267",
    pages = "433--447",
    year = "1986"
}

@article{Awada:1985ep,
    author = "Awada, M. and Townsend, P. K.",
    title = "{$N=4$ Maxwell-einstein Supergravity in Five-dimensions and Its SU(2) Gauging}",
    reportNumber = "Print-85-0266 (CAMBRIDGE)",
    doi = "10.1016/0550-3213(85)90156-7",
    journal = "Nucl. Phys. B",
    volume = "255",
    pages = "617-632",
    year = "1985"
}

@article{Schon:2006kz,
    author = "Schon, Jonas and Weidner, Martin",
    title = "{Gauged N=4 supergravities}",
    eprint = "hep-th/0602024",
    archivePrefix = "arXiv",
    reportNumber = "DESY-06-009, ZMP-HH-06-01",
    doi = "10.1088/1126-6708/2006/05/034",
    journal = "JHEP",
    volume = "05",
    pages = "034",
    year = "2006"
}

@article{Neugebauer:1969wr,
    author = "Neugebauer, G. and Kramer, D.",
    title = "{A method for the construction of stationary einstein-maxwell fields. (in german)}",
    journal = "Annalen Phys.",
    volume = "24",
    pages = "62--71",
    year = "1969"
}

@article{Breitenlohner:1987dg,
    author = "Breitenlohner, Peter and Maison, Dieter and Gibbons, Gary W.",
    title = "{Four-Dimensional Black Holes from Kaluza-Klein Theories}",
    reportNumber = "MPI-PAE-PTH-27-87, LPTENS-87-09",
    doi = "10.1007/BF01217967",
    journal = "Commun. Math. Phys.",
    volume = "120",
    pages = "295",
    year = "1988"
}

@article{Clement:1996nh,
    author = "Clement, Gerard and Galtsov, Dmitri V.",
    title = "{Stationary BPS solutions to dilaton - axion gravity}",
    eprint = "hep-th/9607043",
    archivePrefix = "arXiv",
    reportNumber = "GCR-96-07-02, DTP-MSU-96-11",
    doi = "10.1103/PhysRevD.54.6136",
    journal = "Phys. Rev. D",
    volume = "54",
    pages = "6136--6152",
    year = "1996"
}

@article{Galtsov:1998mhf,
    author = "Gal'tsov, D. V. and Rytchkov, O. A.",
    title = "{Generating branes via sigma models}",
    eprint = "hep-th/9801160",
    archivePrefix = "arXiv",
    reportNumber = "DTP-MSU-98-01",
    doi = "10.1103/PhysRevD.58.122001",
    journal = "Phys. Rev. D",
    volume = "58",
    pages = "122001",
    year = "1998"
}

@article{Tseytlin:1997cs,
    author = "Tseytlin, Arkady A.",
    title = "{Composite BPS configurations of p-branes in ten-dimensions and eleven-dimensions}",
    eprint = "hep-th/9702163",
    archivePrefix = "arXiv",
    reportNumber = "IMPERIAL-TP-96-97-25",
    doi = "10.1088/0264-9381/14/8/009",
    journal = "Class. Quant. Grav.",
    volume = "14",
    pages = "2085--2105",
    year = "1997"
}

@article{Brecher:1999xf,
    author = "Brecher, D. and Perry, M. J.",
    title = "{Ricci flat branes}",
    eprint = "hep-th/9908018",
    archivePrefix = "arXiv",
    reportNumber = "DAMTP-1999-97",
    doi = "10.1016/S0550-3213(99)00659-8",
    journal = "Nucl. Phys. B",
    volume = "566",
    pages = "151--172",
    year = "2000"
}

@article{Erickson:2021psj,
    author = "Erickson, C. W. and Leung, Rahim and Stelle, K. S.",
    title = "{Taxonomy of brane gravity localisations}",
    eprint = "2110.10688",
    archivePrefix = "arXiv",
    primaryClass = "hep-th",
    reportNumber = "Imperial/TP/21/KS/01",
    doi = "10.1007/JHEP01(2022)130",
    journal = "JHEP",
    volume = "01",
    pages = "130",
    year = "2022"
}

@article{Bellorin:2005zc,
    author = "Bellorin, Jorge and Ortin, Tomas",
    title = "{All the supersymmetric configurations of N=4, d=4 supergravity}",
    eprint = "hep-th/0506056",
    archivePrefix = "arXiv",
    reportNumber = "IFT-UAM-CSIC-05-26",
    doi = "10.1016/j.nuclphysb.2005.07.020",
    journal = "Nucl. Phys. B",
    volume = "726",
    pages = "171--209",
    year = "2005"
}

@article{Sen:1992ua,
    author = "Sen, Ashoke",
    title = "{Rotating charged black hole solution in heterotic string theory}",
    eprint = "hep-th/9204046",
    archivePrefix = "arXiv",
    reportNumber = "TIFR-TH-92-20",
    doi = "10.1103/PhysRevLett.69.1006",
    journal = "Phys. Rev. Lett.",
    volume = "69",
    pages = "1006--1009",
    year = "1992"
}

@inproceedings{Sen:1992wi,
    author = "Sen, Ashoke",
    title = "{Black holes and solitons in string theory}",
    booktitle = "{16th Johns Hopkins Workshop on Current Problems in Particle Theory}",
    eprint = "hep-th/9210050",
    archivePrefix = "arXiv",
    reportNumber = "TIFR-TH-92-57",
    pages = "155--174",
    year = "1993"
}

@article{Kallosh:1992ii,
    author = "Kallosh, Renata and Linde, Andrei D. and Ortin, Tomas and Peet, Amanda W. and Van Proeyen, Antoine",
    title = "{Supersymmetry as a cosmic censor}",
    eprint = "hep-th/9205027",
    archivePrefix = "arXiv",
    reportNumber = "SU-ITP-92-13",
    doi = "10.1103/PhysRevD.46.5278",
    journal = "Phys. Rev. D",
    volume = "46",
    pages = "5278--5302",
    year = "1992"
}

@article{Ortin:1992ur,
    author = "Ortin, Tomas",
    title = "{Electric - magnetic duality and supersymmetry in stringy black holes}",
    eprint = "hep-th/9208078",
    archivePrefix = "arXiv",
    reportNumber = "SU-ITP-92-24",
    doi = "10.1103/PhysRevD.47.3136",
    journal = "Phys. Rev. D",
    volume = "47",
    pages = "3136--3143",
    year = "1993"
}

@article{Kallosh:1993yg,
    author = "Kallosh, Renata and Ortin, Tomas",
    title = "{Charge quantization of axion - dilaton black holes}",
    eprint = "hep-th/9302109",
    archivePrefix = "arXiv",
    reportNumber = "SU-ITP-93-3",
    doi = "10.1103/PhysRevD.48.742",
    journal = "Phys. Rev. D",
    volume = "48",
    pages = "742--747",
    year = "1993"
}

@article{Kallosh:1994ba,
    author = "Kallosh, Renata and Kastor, David and Ortin, Tomas and Torma, Tibor",
    title = "{Supersymmetry and stationary solutions in dilaton axion gravity}",
    eprint = "hep-th/9406059",
    archivePrefix = "arXiv",
    reportNumber = "SU-ITP-94-12, UMHEP-407, QMW-PH-94-12",
    doi = "10.1103/PhysRevD.50.6374",
    journal = "Phys. Rev. D",
    volume = "50",
    pages = "6374--6384",
    year = "1994"
}

@article{Clement:2004ii,
    author = "Clement, Gerard and Gal'tsov, Dmitri and Leygnac, Cedric",
    title = "{Black branes on the linear dilaton background}",
    eprint = "hep-th/0412321",
    archivePrefix = "arXiv",
    reportNumber = "DTP-MSU-04-19, LAPTH-1080-04",
    doi = "10.1103/PhysRevD.71.084014",
    journal = "Phys. Rev. D",
    volume = "71",
    pages = "084014",
    year = "2005"
}

@article{Chen:2005uw,
    author = "Chen, Chiang-Mei and Gal'tsov, Dmitri V. and Ohta, Nobuyoshi",
    title = "{Intersecting non-extreme p-branes and linear dilaton background}",
    eprint = "hep-th/0506216",
    archivePrefix = "arXiv",
    reportNumber = "DTP-MSU-05-08, OU-HET-534",
    doi = "10.1103/PhysRevD.72.044029",
    journal = "Phys. Rev. D",
    volume = "72",
    pages = "044029",
    year = "2005"
}

@article{Bergshoeff:1996rn,
    author = "Bergshoeff, E. and de Roo, M. and Eyras, E. and Janssen, B. and van der Schaar, J. P.",
    title = "{Multiple intersections of D-branes and M-branes}",
    eprint = "hep-th/9612095",
    archivePrefix = "arXiv",
    reportNumber = "UG-9-96",
    doi = "10.1016/S0550-3213(97)00151-X",
    journal = "Nucl. Phys. B",
    volume = "494",
    pages = "119--143",
    year = "1997"
}

@article{Tseytlin:1996zb,
    author = "Tseytlin, Arkady A.",
    title = "{On the structure of composite black p-brane configurations and related black holes}",
    eprint = "hep-th/9611111",
    archivePrefix = "arXiv",
    reportNumber = "CERN-TH-96-324, IMPERIAL-TP-96-97-10",
    doi = "10.1016/S0370-2693(97)00037-3",
    journal = "Phys. Lett. B",
    volume = "395",
    pages = "24--27",
    year = "1997"
}

@article{Behrndt:1996pm,
    author = "Behrndt, Klaus and Bergshoeff, Eric and Janssen, Bert",
    title = "{Intersecting d-branes in ten-dimensions and six-dimensions}",
    eprint = "hep-th/9604168",
    archivePrefix = "arXiv",
    reportNumber = "UG-4-96, HUB-EP-96-10",
    doi = "10.1103/PhysRevD.55.3785",
    journal = "Phys. Rev. D",
    volume = "55",
    pages = "3785--3792",
    year = "1997"
}

@article{Lin:2024eqq,
    author = "Lin, Jieming and Skrzypek, Torben and Stelle, K. S.",
    title = "{Compactification on Calabi-Yau threefolds: consistent truncation to pure supergravity}",
    eprint = "2412.00186",
    archivePrefix = "arXiv",
    primaryClass = "hep-th",
    reportNumber = "Imperial/TP/2024/KS/01, DESY-24-179",
    doi = "10.1007/JHEP03(2025)200",
    journal = "JHEP",
    volume = "03",
    pages = "200",
    year = "2025"
}

@article{Cassani:2019vcl,
    author = "Cassani, Davide and Josse, Gr{\'e}goire and Petrini, Michela and Waldram, Daniel",
    title = "{Systematics of consistent truncations from generalised geometry}",
    eprint = "1907.06730",
    archivePrefix = "arXiv",
    primaryClass = "hep-th",
    doi = "10.1007/JHEP11(2019)017",
    journal = "JHEP",
    volume = "11",
    pages = "017",
    year = "2019"
}

@article{Nicolai:1988jb,
    author = "Nicolai, H. and Warner, N. P.",
    title = "{The Structure of $N=16$ Supergravity in Two-dimensions}",
    reportNumber = "DESY-88-129, CERN-TH-5154/88",
    doi = "10.1007/BF01218408",
    journal = "Commun. Math. Phys.",
    volume = "125",
    pages = "369",
    year = "1989"
}

@misc{Figueroa,
author = {Figueroa-O'Farrill, Jos\'e},
title = {Majorana Spinors},
howpublished = {\url{http://www.maths.ed.ac.uk/~jmf/Teaching/Lectures/Majorana.pdf}\,}
}

@article{Maldacena:1997re,
    author = "Maldacena, Juan Martin",
    title = "{The Large $N$ limit of superconformal field theories and supergravity}",
    eprint = "hep-th/9711200",
    archivePrefix = "arXiv",
    reportNumber = "HUTP-97-A097, HUTP-98-A097",
    doi = "10.4310/ATMP.1998.v2.n2.a1",
    journal = "Adv. Theor. Math. Phys.",
    volume = "2",
    pages = "231--252",
    year = "1998"
}

@article{Witten:1998qj,
    author = "Witten, Edward",
    title = "{Anti de Sitter space and holography}",
    eprint = "hep-th/9802150",
    archivePrefix = "arXiv",
    reportNumber = "IASSNS-HEP-98-15",
    doi = "10.4310/ATMP.1998.v2.n2.a2",
    journal = "Adv. Theor. Math. Phys.",
    volume = "2",
    pages = "253--291",
    year = "1998"
}

@article{Coimbra:2012af,
    author = "Coimbra, Andre and Strickland-Constable, Charles and Waldram, Daniel",
    title = "{Supergravity as Generalised Geometry II: $E_{d(d)} \times \mathbb{R}^+$ and M theory}",
    eprint = "1212.1586",
    archivePrefix = "arXiv",
    primaryClass = "hep-th",
    reportNumber = "IMPERIAL-TP-12-DW-01",
    doi = "10.1007/JHEP03(2014)019",
    journal = "JHEP",
    volume = "03",
    pages = "019",
    year = "2014"
}

@article{Strominger:1996sh,
    author = "Strominger, Andrew and Vafa, Cumrun",
    title = "{Microscopic origin of the Bekenstein-Hawking entropy}",
    eprint = "hep-th/9601029",
    archivePrefix = "arXiv",
    reportNumber = "HUTP-96-A002, RU-96-01",
    doi = "10.1016/0370-2693(96)00345-0",
    journal = "Phys. Lett. B",
    volume = "379",
    pages = "99--104",
    year = "1996"
}

\end{document}